\documentclass[aps,prb,preprint]{revtex4}
\usepackage{graphicx,amsmath}
\usepackage{epstopdf}
\begin{document}
\title{Valence band offset, strain and shape effects on confined states in self-assembled InAs/InP and InAs/GaAs quantum dots}
\author{M. Zieli\'nski}
\email{mzielin@fizyka.umk.pl}
\affiliation{Institute of Physics, Faculty of Physics, Astronomy and Informatics, Nicolaus Copernicus University, Grudziadzka 5, 87-100 Torun, Poland}
\begin{abstract}
I present a systematic study of self-assembled InAs/InP and InAs/GaAs quantum dots single particle and many body properties as a function of quantum dot-surrounding matrix valence band offset.
I use an atomistic, empirical tight-binding approach and perform numerically demanding calculations for half-million atom nanosystems.
I demonstrate that the overall confinement in quantum dots is a nontrivial interplay of two key factors: strain effects and the valence band offset.
I show that strain effects determine both the peculiar structure of confined hole states of lens type InAs/GaAs quantum dots and
the characteristic ``shell-like'' structure of confined holes states in commonly considered ``low-strain'' lens type InAs/InP quantum dot.
I also demonstrate that strain leads to single band-like behavior of hole states of disk type (``indium flushed'') InAs/GaAs and InAs/InP quantum dots.
I show how strain and valence band offset affect quantum dot many-body properties: the excitonic fine structure,
an important factor for efficient entangled photon pair generation, and the biexciton and charged excitons binding energies.
\end{abstract}
\maketitle

\section{Introduction}
Fully ab-initio, parameters free, modeling of million atom self-assembled\cite{arek-book} or nanowire\cite{bjork} quantum dots is still beyond the
reach of current computers. For practical, atomistic calculation, semi-empirical approaches like the empirical tight-binding (ETB)
\cite{schulz-schum-czycholl,lee-johnsson-prb2001,leung-whaley-prb1997,diaz-bryant,usman1,usman2,santoprete,jaskolski-zielinski-prb06,korkusinski-zielinski-hawrylak-jap09,zielinski-prb09,sheng-cheng-prb2005,zielinski-including,bryant-fss,bryant-fss2}
or the empirical pseudopotential method (EPM)\cite{wang-zunger,wang-will-zunger-siang-singh,canning-zunger-jcompphys2000,bester-zunger-prb2005,Zunger-EPM1,Zunger-EPM2,gong-prb77,gong-complexes} are typically employed.
The computation scheme usually starts with strain field calculation followed by the single particle calculation followed then with
the configuration interaction approach to obtain many-body (exciton, charged exciton, multi-exciton) spectra.\cite{zielinski-prb09,sheng-cheng-prb2005}

Semi-empirical approaches use sets of fitted parameters determined to reproduce bulk properties like effective masses, bulk deformation potentials and gaps at different points of the Brillouin zone\cite{jancu}.
Bulk derived parameters are later used for calculation of nanosize systems. Apart from potential (``bulk to nanosystem'') transferability issues, one may question the reliability
of important parameters describing the bulk electronic structure,
that act as the input data for the empirical fitting procedure.
For example, in the case of the InAs absolute valence band deformation potential ($a_v$) not even the sign of this quantity is unambiguously determined.
\cite{VBO-Vurgraftman,Wei-DFT2,Wei-Zunger-DFT,VanDeWalle,kend-hart-zunger,kadantsev-ziel-kork-haw,kadantsev-ziel-haw,kadantsev}
As semiconductor (self-assembled or nanowire) quantum dots typically are mixed material systems an additional empirical bulk parameter, the valence band offset between quantum dot and surrounding matrix material, has to be incorporated into the Hamiltonian. This ``natural''\cite{li-natural-wei,kadantsev} valence band offset (VBO) determines the depth of unstrained hole and electron confining potentials and combined with strain and deformation potentials constitutes the overall confining potential for the strained case, i.e. the ``strained'' band offset.
There is again a substantial uncertainty of the natural VBO values, e.g., the reported InAs/GaAs VBO varies from $50$ to $500$~meV.\cite{VBO-Vurgraftman,li-natural-wei}
The source of this discrepancy is not only due to the difference between experimental and theoretically reported values, but ``It is to be emphasized, however,
that even within the framework of Kohn-Sham DFT, different computational schemes result in different predictions for the natural band offsets [...]''.\cite{kadantsev}
Y.H. Li and co-workers\cite{li-natural-wei} state additionally that there has been ``long-standing anomalies between theory and experiment'' and in their ab-initio calculation
``For GaAs/InAs the predicted offset is increased from $0.06$~eV in the previous calculation to $0.50$~eV''.

For a given pair of materials (e.g. InAs/GaAs) the VBO has only one value, however as discussed above, typically this value is practically unknown or given with large ($>100$~meV) uncertainty.
For example, the VBO~$=0.06$~eV is used in the EPM approach, the VBO~$\approx0.17$~eV is utilized in {\bf k$\cdot$p} studies,\cite{kp-schliwa} the VBO~$=0.23$~eV is incorporated in the ETB model by Boykin et al.\cite{boykin-diagonal-shift}, whereas VBO~$\approx500$~meV is reported by the DFT calculations.\cite{li-natural-wei}
In this paper, as a practical resolution of this problem, I utilize an approach in which I perform calculations by an artificial variation of the VBO over wide range of values.
While the VBO has been considered as a merely technological parameter of lesser importance, the current study shows that it is quite the opposite.
Both the single particle and the many body properties are affected by the choice of the VBO and the caution should be exercised before applying different VBO values in a semi-empirical calculation.
Abstracting from the experimental reality (there is only one, yet unknown VBO value, for a given pair of materials), the ``artificial'' modification of the VBO is by itself a very interesting theoretical
tool to study the effects of band confinement versus other effects such as strain. For example, the calculation of spectra for InAs/InP and InAs/GaAs lens type quantum dots using the same (nonetheless artificial)
VBO helps to understand the difference between both types of nanosystems.


The ambiguity of the $a_v$ and VBO bulk material values may thus affect\cite{zielinski-including} the accuracy of the qualitative description of confined
valence band states in semiconductor nanosystems. This is further important as recent empirical pseudopotential method papers (e.g. Refs. \cite{gong-prb77} or \cite{singh-bester-eh} etc.) still refer to older EPM parameterizations\cite{Zunger-EPM2} utilizing the ``questioned''\cite{li-natural-wei} VBO value $\approx 50$~meV. The Section E of the current paper discusses this point showing that whereas ``natural'' VBOs in ETB and EPM
can differ significantly, strained band offsets in the two approaches are very similar.

Despite over a decade of intensive studies the problem of a detailed understanding of holes in self-assembled InAs quantum dots is still an active field of research\cite{blokland}.
Ediger et al. \cite{ediger-zunger-aufbau} observed characteristic spectral structure of hole states in InAs/GaAs lens type quantum dots leading to a non trivial hole charging pattern of excitonic complexes. In our earlier work\cite{korkusinski-zielinski-hawrylak-jap09} we have noticed that InAs/GaAs quantum dot shape affects spectral properties of holes significantly.
Recently Gong et al.\cite{gong-prb77,gong-complexes} have calculated electronic structure of InAs/InP lens type quantum dots and speculated that differences with respect to analogous InAs/GaAs systems are due to different common ions or different VBOs in both systems.
In this paper, by a systematic VBO analysis, I show it is not the VBO, but rather strain effects that play a dominant role in determining the character of single particle holes states in quantum dots.

A two photon cascade from the quantum dot biexciton state can generate entangled photons\cite{benson-fss} and has attracted a great interest for applications in quantum information.
However, anisotropic exchange splitting of bright excitons\cite{bayer-eh}, induced by the asymmetry of quantum dot confining potential\cite{singh-bester-eh}, inhibits entanglement.
Understanding the origins of fine structure splitting is thus of great importance for potential quantum dot applications.
Recently, yet another scheme for entangled photons generation has been proposed\cite{avron-fss,ding-fss,reimer-fss} based on tuning the biexciton binding energy to zero.
As in a typical experiment different excitonic complexes (both charged and neutral) are observed together, the prediction of the spectral line order\cite{gong-complexes} or binding energies (such as that of the biexciton)
is usually far from trivial.
To solve such issues, an inverse approach for quantum dot calculation has been recently proposed\cite{mlinar-loop}.
In this method one uses excitonic spectroscopy experimental data to determine quantum dot structural properties.
However the accuracy of such prediction must depend on the accuracy of the many-body calculations and indirectly on empirical parameters (such a the VBO) used in a calculation.

In this paper I compare properties of strained and unstrained systems, study the role of quantum dot shape and the evolution of single particle energies and charge probability densities as a function of the valence band offset.
By a systematic VBO analysis, I show that strain and valence band offset effects play different, important roles in determining the character of single particle states in quantum dots.
Finally, I show that the choice of the VBO affects substantially many body energies, in particular biexciton and trions binding energies and the excitonic fine structure.

\section{Systems and methods}
In the following I present a systematic study of lens and disk type InAs quantum dots, surrounded by either InP or GaAs matrix, as a function of the valence band offset, with strain effects included or artificially neglected. The height of the disk type quantum dot is $h=3$ nm and the base diameter is $D=16.8$ nm. For the sake of comparison with the EPM calculations\cite{Zunger-EPM1,Zunger-EPM2,gong-prb77,gong-complexes} the height of the lens type dot is chosen as $h=3.5$~nm and the base diameter is $D=25$~nm. Both dots are located on a $1$~nm thick wetting layer.
The presence of the wetting layer is particularly important for disk type quantum dots as it lowers the overall quantum dot symmetry from $D_{2d}$ to $C_{2v}$ (the lack of ``rotoinversion operation''\cite{singh-bester-eh}) and therefore both kinds (lens and disk) of quantum dots have low $C_{2v}$ symmetry.

The calculation consists of several major steps: first atomic positions are calculated.
There is a lattice mismatch between the quantum dot material (InAs) and the surrounding matrix material (GaAs or InP).
To calculate strain relaxed positions I use the atomistic valence force field (VFF) approach of Keating\cite{keating}.
This method is described in more detail in Refs.\cite{pryor-zunger,saito-arakawa} and in our previous works\cite{jaskolski-zielinski-prb06,zielinski-prb09}.
The size of the computational domain, including more than $50$ millon atoms, guarantees convergence of the strain distribution\cite{zielinski-multiscale}.

Due to the small lattice mismatch of InAs and InP I neglect the piezoelectric effects in the present calculation, following similar arguments by Gong et al.\cite{gong-prb77} who ignore piezoelectricity in the empirical pseudopotential work on InAs/InP quantum dots. Consistently, piezoelectric effects can also be neglected for low aspect ratio\cite{kp-schliwa,usman1,usman2} lens and disk type InAs/GaAs quantum dots,
where the piezoelectricity is either negligible\cite{bester-zunger-prb2005} or the contribution due to second-order effects tends to cancel linear terms\cite{bester-zunger-wu-vanderbilt-prb2006,kp-schliwa}.
In particular for lens type InAs/GaAs Ref.\cite{kp-schliwa} states that: ``for smaller aspect ratios [...] first- and second-order effects compensate each other with respect to their impact on the electronic states.''
Finally the piezoelectricity is neglected for a sake of a fair comparison with other approaches where this effect is neglected (comment 41 from Ref.\cite{Zunger-EPM1}).

In the second step of the calculation, the single particle states are obtained by building the $sp^3d^5s^*$ tight-binding Hamiltonian\cite{jancu,zielinski-including} and then diagonalizing
the Hamiltonian by means of the Arnoldi algorithm with the matrix-vector multiplication parallelized using {\em OpenMP} approach on $48$ core, shared memory system.

The single-particle tight-binding Hamiltonian for the system of $N$ atoms and $m$ orbitals per atom can be written in the language of the
second quantization (in the site basis) in the following form:
\begin{equation}
  \hat{H}_{TB} =
  \sum_{i=1}^N \sum_{\alpha=1}^{m}
       E_{i\alpha}c_{i\alpha}^+c_{i\alpha}
+  \sum_{i=1}^N \sum_{\alpha=1,\beta=1}^{m}
       \lambda_{i\alpha,\beta}c_{i\alpha}^+c_{i\beta}
+   \sum_{i=1}^N \sum_{j=1}^{4} \sum_{\alpha,\beta=1}^{m}
       t_{i\alpha,j\beta}c_{i\alpha}^+c_{j\beta}
\end{equation}
where $c_{i\alpha}^+$ ($c_{i\alpha}$) is the creation (annihilation)
operator of a carrier on the orbital $\alpha$ localized on the site
$i$, $E_{i\alpha}$ is the corresponding on-site (diagonal) energy, and
$t_{i\alpha,j\beta}$ describes the hopping (off-site, off-diagonal)
of the particle between the orbitals on (4) nearest neighboring sites.
Coupling to further neighbors is neglected.
Finally, $\lambda_{i\alpha,\beta}$ (on-site, off-diagonal)
accounts for the spin-orbit interaction
following the description given by Chadi \cite{chadi-so-in-tb}.

As the quantum dot/matrix material lattice constants do not enter TB Hamiltonian explicitly, strain in the TB method is accounted by the modification of Hamiltonian matrix elements from the bulk (unstrained) values to the values modified due to bond lengths/angles modification. Therefore if one uses InAs/(GaAs,InP) bulk Hamiltonian matrix elements, one is simply neglecting strain effects.
Therefore for the (artificially) unstrained\cite{jaskolski-zielinski-prb06} systems I use bulk TB parameters set from Ref.\cite{jancu} and thus there is no strain contribution in the Hamiltonian nor the relaxation of atomic position is accounted for. In other words the ``unstrained'' system (``strain effects neglected'') corresponds to strain ``unrelaxed'' system with the VFF step and the modification of TB parameters neglected.
For strained systems, since strain effects change bond lengths and angles, strain relaxed positions are used to modify TB parameters (diagonal and off-diagonal matrix elements) following the description given in detail in my earlier work\cite{zielinski-including}.

I have calculated the single particle spectra for many VBO values from $50$ meV to $500$ meV with a $10$ meV step. 
In this paper I use a multi-scale (multi-domain) approach where smaller computational domain is used for the single particle calculation\cite{zielinski-multiscale,lee-boundary}.
Yet, as the number of atoms in the TB domain is larger than half a million, in order to make the entire process feasible I calculate only eigenenergies of several lowest electron and holes confined states. Then for several chosen VBO values I additionally calculate eigenstates and plot corresponding probability charge densities.

Finally, for several VBO values electron and hole Coulomb matrix elements (Coulomb and exchange integrals) are calculated according to the approach given in Ref.~\cite{zielinski-prb09}.
In a GW approach\cite{onida} one calculates the effective interaction $W$ self-consistently.
Not being able to carry out this calculation, I assume a statically screened Coulomb interaction.
Hence the Coulomb matrix elements $V_{ijkl}$ are given by:
\begin{equation}
V_{ijkl}=\int \int \phi_i^* \left(\vec{r_1}\right) \phi_j^* \left(\vec{r_2}\right)
\frac{e^2}{\epsilon\left(\vec{r_1},\vec{r_2}\right)\left|\vec{r_1}-\vec{r_2}\right|}
\phi_k \left(\vec{r_2}\right) \phi_l \left(\vec{r_l}\right)
\label{coulomb-general}
\end{equation}
where $\epsilon\left(\vec{r_1},\vec{r_2}\right)$ is the position-dependent dielectric function and $\phi$ are single-particle wave functions.
By substituting single-particle wave functions in the form of linear combination of atomic orbitals: $\phi_i=\sum_{\vec{R},\alpha}b^i_{\vec{R}\alpha}|\vec{R}\alpha\rangle$ into Eq.~\ref{coulomb-general} and then by utilizing a series of approximations\cite{schulz-schum-czycholl,zielinski-prb09} (including the two-center approximation and retaining monopole-monopole contributions only) one obtains an approximate form of Coulomb matrix elements\cite{zielinski-prb09}:
\begin{eqnarray}
V_{ijkl}=
\sum_{\vec{R_1}}\sum_{\vec{R_2}\neq\vec{R_1}}
\left[\sum_{\alpha_1}
b_{\vec{R_1}\alpha_1}^{i*}b_{\vec{R_1}\alpha_1}^{l}\right]
\left[\sum_{\alpha_2}
b_{\vec{R_2}\alpha_2}^{j*}b_{\vec{R_2}\alpha_2}^{k}\right]
\frac{e^2}{\epsilon\left|\vec{R_1}-\vec{R_2}\right|}
+\nonumber \\
\sum_{\vec{R_1}}
\sum_{\alpha_1\alpha_2\alpha_3\alpha_4}
b_{\vec{R_1}\alpha_1}^{i*}b_{\vec{R_1}\alpha_2}^{j*}
b_{\vec{R_1}\alpha_3}^{k}b_{\vec{R_1}\alpha_4}^{l}
\langle\vec{R_1}\alpha_1,\vec{R_1}\alpha_2
\left|\frac{e^2}{\left|\vec{r_1}-\vec{r_2}\right|}\right|
\vec{R_1}\alpha_3,\vec{R_1}\alpha_4
\rangle.
\label{coulomb-special}
\end{eqnarray}
The first term is the long-range, bulk-screened, contribution to the two-center integral built from the monopole-monopole interaction\cite{Franceschetti,Goupalov} of two charge densities localized at different atomic sites.
The second term is the on-site unscreened part, calculated by direct integration using atomic (Slater) orbitals\cite{lee-johnsson-prb2001,leung-whaley-prb1997}.
This approach is justified by the fact that the screening (Thomas-Fermi) radius ($\approx2-4$\AA) is on the order of bond length \cite{lee-boundary,lee-johnsson-prb2001} resulting in nearly bulk screening of off-site (long-range) terms and limited screening of on-site (short-range) terms contribution.

As noticed by Leung and Whaley~\cite{leung-whaley-prb1997}: ``the precise magnitudes of on-site integrals may depend on the orbital basis employed to evaluate them''.
Lee and coworkers~\cite{lee-johnsson-prb2001} compared the results obtained with Slater-type orbitals and Gaussian-type orbitals and also studied the role of basis orthogonality.
They concluded that the use of nonorthogonal Slater orbitals can be estimated to imply about $20\%$ overall uncertainty in the on-site integrals and that tight-binding descriptions of electron-hole Coulomb interactions in quantum dots should be reliable for quantum dots larger than about $0.20$~nm, thus much smaller than quantum dots studied in this paper.
Ref.~\cite{lee-johnsson-prb2001} also states that the sensitivity of the basis orbitals decreases quickly as the dot size increases.
On the other hand, Franceschetti and coworkers\cite{Franceschetti} shown that not only Coulomb interaction, but also exchange interaction is dominated by the long-range component, and the short-range component constitutes only of about $20\%$ of the total excitonic exchange interaction (exchange splitting of the lowest exciton states), whereas the monopole-monopole contributions capture up to $90\%$ of the long-range component for typical III-V systems (Table I from Ref.\cite{Franceschetti}).
This conclusion is further supported by Luo and coworkers\cite{Franceschetti2} showing that in general direct-gap quantum dots (such as InAs) the electron-hole exchange interaction is dominated by the long-range
component and the (small) short range contribution scales as $1/R^3$ ($R$ being the dot radius).
In the current paper, consistent with above discussions, the short-range, and thus basis depended, contribution accounts for only about $1\%$ of the ground electron-hole states Coulomb direct attraction, about $20\%$ of the excitonic exchange splitting (dark-bright exciton splitting\cite{bayer-eh}) and about $10\%$ of the excitonic fine structure (``anisotropic exchange''\cite{bayer-eh}) splitting (both for bright and dark excitons splitting).
The upper bound for the basis uncertainty error can thus be expected not to exceed the above fractions and typically should be significantly lower.

The Hamiltonian for the interacting electrons and holes can be written in second quantization as:\cite{hawrylak_korkusinski_topicsapplphys2003}
\begin{eqnarray}
  \hat{H}_{ex} =  \sum_{i}E_i^ec_i^\dagger c_i+\sum_{i}E_i^hh_i^\dagger h_i
  +\frac{1}{2}\sum_{ijkl}V_{ijkl}^{ee} c_i^\dagger c_j^\dagger c_k c_l
  +\frac{1}{2}\sum_{ijkl}V_{ijkl}^{hh} h_i^\dagger h_j^\dagger h_k h_l \nonumber \\
  -\sum_{ijkl}V_{ijkl}^{eh,dir} c_i^\dagger h_j^\dagger h_k c_l
  +\sum_{ijkl}V_{ijkl}^{eh,exchg} c_i^\dagger h_j^\dagger c_k h_l
\end{eqnarray}
The many-body Hamiltonian for the exciton (X), the biexciton (XX), positively ($X^-$) and negatively charged ($X^+$) trions is solved using the configuration interaction approach\cite{zielinski-prb09,sheng-cheng-prb2005}.

The quantum dot and the surrounding matrix may share the same anion (e.g. InAs/GaAs), cation (e.g. InAs/InP) or not have a common ion (e.g. InAs/GaP).
In the empirical tight-binding the treatment of quantum dot/material interface atoms is ambiguous.
In Boykin et al.\cite{boykin-diagonal-shift} approach, this is handled during the fitting procedure where the diagonal Hamiltonian matrix elements of the common atom are kept the same in both materials. The value on the material band offset is incorporated into off-diagonal hopping matrix elements.
This approach removes the necessity of modifying on-site matrix elements for interface atoms, but complicates significantly for ternary systems like InAs/InP/GaP, where, e.g., the bulk GaP properties are indirectly coupled to InAs bulk properties through the fitting process.
Additionally as VBOs are embedded into the tight-binding parameterization, a necessary, complicated refit would be needed for every different VBO value.
In this paper I use an approach\cite{zielinski-including} in which I account for the valence band offset by shifting diagonal matrix elements of the quantum dot material.
On-site matrix elements of interface atoms are calculated as a weighted sum of neighboring atoms materials on-site parameters.
The Hamiltonian is built and then diagonalized for each VBO value: this gives a capability of freely tuning the VBO, suitable thus for the VBO dependence studies.

\section{Results}
Figure~\ref{lens-example} shows the evolution of single particle electron and hole energy levels for the lens type InAs/GaAs quantum dot as a function of the InAs/GaAs VBO.
With the increasing VBO value there is an energy upshift due to overall ``reference level'' (VBO) energy shift, whereas the effective band gap $E_{gap}=e_1-h_1$ does not change considerably\cite{santoprete,jaskolski-zielinski-prb06}.
In this paper I show that whereas the above statement is generally true\cite{santoprete,jaskolski-zielinski-prb06,niquet-band-offset}, the situation is far more complicated for spectral quantities other $E_{gap}$.
To analyze the specifies of the electron and the hole spectra in the following plots I subtract the corresponding carrier ground state energy as illustrated
on Figure~\ref{lens-example}(b) and~\ref{lens-example}(c).

Valence band offset is related to (``natural'' or unstrained) conduction band offset CBO through the following relation $CBO=matrix_{gap}-dot_{gap}-VBO$, where $matrix_{gap}$ is the bulk band gap of the matrix material (GaAs or InP) and $dot_{gap}$ is the bulk band gap of the quantum dot (InAs) material.
For the VBO in $50-500$~meV range, the CBO varies correspondingly from $\approx1$ to $\approx0.5$~eV, with little difference between InAs/GaAs and InAs/InP cases, due to $0.1$ eV difference of InP and GaAs bulk band gaps.
Increasing the VBO corresponds thus to decreasing of the CBO, i.e. lower confinement of the electron states.
Even though CBO is nominally larger than VBO, effects of CBO variation on electron states should be observable due to their lower confinement (effective mass).

Figure~\ref{disc-en} and \ref{lens-en} show energy levels corresponding to several lowest electron (upper-$CB$) and hole (lower-$VB$) states calculated for InAs/InP and InAs/GaAs disk type and lens type quantum dots as function of quantum dot-matrix valence band offset (VBO) with strain effects either artificially neglected or included.
Figures from Figure~\ref{disc-el} to Figure~\ref{lens-ho-str} show corresponding charge/probability density isosurfaces.
These figures contain substantial amount of information and will be analyzed in detail in the following part of the text.

\subsection{Electron states - strain effects neglected}
In an artificially unstrained InAs/(InP,GaAs) disk type quantum dot energy spectra of lowest electron levels reveal shell-like structure [Fig.~\ref{disc-en} (a) and (b)], with the ground electron state of \textit{s}-type character [Fig.~\ref{disc-el}].
Despite the absence of strain and the rotational shape symmetry of the disc quantum dot, the presence of atomistic interfaces and low symmetry of underlying crystal lattice introduces the asymmetry into the Hamiltonian\cite{bester-zunger-prb2005}. Thus, there are two closely spaced (splitting $<1.5$meV) excited states ($e_2$ and $e_3$) of \textit{p}-like character (of approximate angular momentum character $L=\pm1$), followed by two excited ($e_4$ and $e_5$) closely spaced states (splitting $\approx 4$ meV) of \textit{d}-like character. Splittings within \textit{p}- and \textit{d}-shell are however much smaller than spacings between different shells: \textit{s}-\textit{p} ($60$ meV) and \textit{p}-\textit{d} ($70$ meV).
The $e_6$ state (of ``\textit{2s}'' character \cite{bester-zunger-prb2005}) is separated from the lower lying \textit{d}-like states $e_4,e_5$ by $\approx 20$ meV, a hallmark of disc-like confinement\cite{arek-book}.
Such structure of levels with quasi-degenerate energies corresponding to $L=0,\pm1,\pm2,...$ is to be expected for nominally cylindrical disc-shaped quantum dots.

Charge probability densities corresponding to several lowest electron states in an artificially unstrained InAs disk type quantum dot practically do not change in the considered range of VBO values and are very similar for both InP and GaAs matrices [Fig.~\ref{disc-el}].
For small VBO~$\approx 100$~meV there is a slight elongation of the \textit{p}-shell states ($e_2$ and $e_3$) along $[110]$ and $[1\underline{1}0]$ crystal axis, but otherwise these states have well defined cylindrical-like symmetry.
For the VBO changing from $0.1$ to $0.4$~eV, the corresponding ground electron state localization inside a quantum dot drops only by $3$\%, i.e. from $81$\% to $78$\% in the InAs/InP case and from $86$\% to $83$\% in the InAs/GaAs case.
With no strain effects included, the confinement is generally somewhat lower for InAs/InP systems when compared to InAs/GaAs due to lower InP band gap (and thus lower $CBO$).

For the disk type quantum dot, in the absence of strain, spacings between different shells (\textit{s}-\textit{p} and \textit{p}-\textit{d}) do not change much as a function of the VBO, however Figure~\ref{p-shell} shows that the splitting of the electron \textit{p}-shell (``\textit{p}-shell anisotropy'') increases monotonically (quasi-parabolically) with decreasing confinement, most likely due to the increasing role of interface effects in a progressively shallower (decreasing CBO) well for electrons.
Notably, in the absence of strain, the \textit{p}-shell splitting is systematically larger in the disk type InAs/InP quantum dot than in the disk type InAs/GaAs system, which I speculate, can also be related to ($\approx 5$\%) lower confinement of electrons in InAs/InP quantum dots.
Another important difference between InAs/GaAs and InAs/InP systems is that in the first case the quantum dot and the surrounding material share common anion (As), whereas in the latter case they share common cation (In). Since electron wave functions are more localized on cation sites, having quantum dot and matrix material with same cations will increase the amplitude of the electron wave function at the interface, thus may increase the \textit{p}-shell splitting for InAs/InP systems when strain effects are neglected.

In the effective mass approximation, lens type quantum dots are expected to show 2D harmonic oscillator-like spectrum \cite{arek-book,wojs-hawrylak}.
In my atomistic calculations for the lens type quantum dot [Fig.~\ref{lens-en} (a) and (b)] I also observe the well pronounced shell structure of the electron levels and the well defined nodal-structure of corresponding charge densities (Fig.~\ref{lens-el}).
For the lens type quantum dot, electron \textit{s}-\textit{p} and \textit{p}-\textit{d} level spacings slightly decrease with the increasing VBO.
The ground electron state charge distribution is apparently not affected by the choice of the matrix material.
The lower electron \textit{p}-shell ($e_2$) state is localized along [1\underline{1}0] crystal axis and the higher \textit{p}-shell state is localized along [110] axis in the InAs/GaAs system, whereas in the InAs/InP system both \textit{p}-shell states maintain cylindrical-like symmetry.
Figure~\ref{p-shell} shows that with strain effects neglected the \textit{p}-shell splitting depends more significantly on the absolute depth of the confining potential due to the VBO rather than on the particular quantum dot shape.
Higher lying states are approximately of the same symmetry for both GaAs and InP matrices (Fig.~\ref{lens-el}) with similar inter-shell spacings ($\approx 60$~meV).

\subsection{Electron states - strain effects included}
With strain effects accounted for, the first important difference is a significant increase of the \textit{p}-shell splitting for both disc and lens system (Fig.~\ref{p-shell}) and the well pronounced elongation of the \textit{p}-shell states along [1\underline{1}0] and [110] crystals axes (Fig.~\ref{disc-el}).
Interestingly the \textit{p}-shell states elongation for the disk type quantum dots is opposite to that of the lens type quantum dots (Fig.~\ref{lens-el}).
Also, for the lens type quantum dot the orientation of the \textit{p}-shell states is reversed\cite{bester-zunger-prb2005} when compared to the strain-free case:
the anisotropy due to atomic interface is thus reversed by the anisotropy due to strain.
For the highly strained InAs/GaAs disk type quantum dot, the competition of two different anisotropy sources manifest itself by a non-monotonic change of the \textit{p}-shell splitting as a function of the VBO (Fig.~\ref{p-shell}).
With strain effects included the \textit{p}-shell splitting is generally higher in lens type than in disk type quantum dots, most likely due to curved quantum dot shape or larger surface/volume ratio in lens type systems.
With strain effects neglected the \textit{p}-shell splitting was generally larger for InAs/InP systems, however with strain effects included this trend reverses, the splitting is dominated by strain, and therefore is larger in InAs/GaAs than in InAs/InP systems.
It is interesting to notice that whereas electron states are built predominately from \textit{s}-type atomic orbitals (and thus should be affected predominantly by the non-directional hydrostatic strain) one expects non-zero electron-hole coupling\cite{kane} and thus one can speculate that the effects of (biaxial) strain on hole states (which will be discussed later) may also indirectly affect the electron \textit{p}-shell splitting.

To summarize, the overall structure of \textit{p}-shell electron states is determined by combining: the matrix material, the quantum dot shape and the VBO value and none of these factors can be neglected
\footnote{Incorporation of piezoelectricity, necessary for high aspect ratio (``tall'') quantum dots, would further complicate that picture.}.
For confined quantum dot states (both electrons and holes) spacings between different shells (\textit{s}-\textit{p}, \textit{p}-\textit{d}, etc.) increase with the increasing confining potential depth (i.e. band offset, the CBO and the VBO for electron and holes correspondingly) as expected from the quantum confinement effect [Fig.~\ref{disc-en} and Fig.~\ref{lens-en}].
However splitting of levels within a given shell decreases with the increasing confinement. This is due the progressively larger localization and effectively smaller influence of the material interface, which acts as the source of splitting.
Consequently, higher lying (\textit{d}-shell) states properties are even more susceptible to the choice of the VBO due to their lower confinement.

\subsection{Hole states - strain effects neglected}
Interestingly even when strain effects are neglected, the ground hole state of the considered disc and lens type quantum dots has well defined \textit{s}-like symmetry [Fig.~\ref{disc-ho-nstr} and Fig.~\ref{lens-ho-nstr}] and is predominately of heavy-hole character.
It initially may sound surprising as in this case there are no strain related heavy hole-light hole splitting terms in the Hamiltonian\cite{bir-pikus}.
However, the quasi-two-dimensional confinement in flat quantum dot systems is efficient enough to separate both type of holes.
Alternatively, one can associate heavy hole states with the in-plane component dominated by $p_x$ and $p_y$ atomic orbitals\cite{jaskolski-zielinski-prb06,yu-cardona} and affected by small, lateral confinement.
Then the light-hole states are the predominantly constituted by $p_z$ orbitals and highly influenced by the vertical confinement,
and thus energetically shifted away from the ground hole state.

Higher lying states however, for both types of quantum dots, reveal complex, mixed angular momentum character [Fig.~\ref{disc-ho-nstr} and Fig.~\ref{lens-ho-nstr}] and show no clear shell-like structure of their energy spectra [Fig.~\ref{disc-en} and Fig.~\ref{lens-en}],
that was so characteristic for the single band-like confinement of the electron states.
For lens type quantum dots, the first and the second excited hole states are not even of the \textit{p}-like symmetry and no nodal planes are observed (Fig.~\ref{lens-ho-nstr}).

Due to strong ($>90\%$) confinement of hole states in quantum dot area, with strain effects neglected, there is little difference between hole states properties with respect to the surrounding matrix (GaAs/InP). This is particularly true for large VBO values, and better confined states, where the surrounding matrix (interface effects) plays a lesser role. Consequently there is almost one to one correspondence of charge densities for the disk type InAs/InP and InAs/GaAs quantum dot cases (Fig.~\ref{disc-ho-nstr}).

Interestingly, even for the VBO$=0$~eV and no strain effects included I still observe quantum dot hole confined states due to material properties (``effective mass'') discontinuity (Fig.~\ref{disc-ho-nstr}).

\subsection{Hole states - strain effects included}
Strain affects holes states significantly.
For the disk type InAs/GaAs and InAs/InP quantum dots, in the ``realistic''\footnote{Whereas the reported InAs/GaAs VBO extreme values vary from $60$~meV to $500$~meV,
the more ``realistic'' or the ``recommended''\cite{VBO-Vurgraftman} values can be limited in the more narrow range.} range of VBO values ($210-350$~meV),
the structure of confined hole states is in a vivid contrast to the strain-free case and resembles that of single band electron confined states,
with well pronounced shells of \textit{p}-like and \textit{d}-like symmetry.
This effect is well visible, both in the charge distributions [Fig.~\ref{disc-ho-str}] and in the energy spectra [Fig.~\ref{disc-en} (g) and (h)].
The hole \textit{p}-shell splitting lies within few meV, i.e. much smaller than ($25-30$~meV) \textit{s}-\textit{p} level spacing.
Strain lifts heavy hole-light hole degeneracy\cite{bir-pikus,pryor-zunger,yu-cardona} and thus effectively decouples light-hole component of the confined hole function, leading to the single band-like behavior of hole states in disk type (``indium flushed''\cite{wasilewski-fafard-maccaffrey}) quantum dots and resulting in the characteristic shell-structure know also from the experiment.\cite{raymond-studenikin-prl2004} Similar conclusion may also be drawn for the disc-shaped nanowire quantum dots (which are however not placed on the wetting layer).
Whereas we have obtained this characteristic spectrum in our earlier work\cite{korkusinski-zielinski-hawrylak-jap09}, we have attributed it to the quantum dot shape. In this paper, I demonstrate that strain, rather than shape only, is responsible for the typical spectra of disk type quantum dots.
Additionally, strain actually leads also to the characteristic harmonic oscillator-like structure of confined holes states
in lens type InAs/InP quantum dot as seen in Fig.~\ref{lens-en} (g) and Fig.~\ref{lens-ho-str}.
In this case no shell-like structure could be observed with strain effects artificially neglected [Fig.~\ref{lens-en} (e) and Fig.~\ref{lens-ho-nstr}].
Therefore strain cannot be neglected even for InAs/InP quantum dots, commonly considered ``low-strain'' systems.

Holes states properties may vary significantly in the considered range of the VBO values, as VBO~$=0$ corresponds to no confinement other than due to strain and material properties (``effective mass'') discontinuity (Fig.~\ref{disc-ho-str}).
For small VBO values (and lesser confinement) there is a significant leakage of the hole wavefunction into the surrounding matrix and into the highly biaxially strained wetting layer
[Fig.~\ref{disc-ho-str} and Fig.~\ref{lens-ho-str}].
Strained disk type quantum dots reveal thus a strong dependence on the choice of the valence band offset, however for VBO~$>300$~meV their spectral properties stabilize.

Higher lying holes states have mixed angular momentum character and their evolution with respect to the VBO seems to depend in a complicated way on the relative
evolution of different angular momenta components. In terms of states localization I can label hole states by those localized predominately
along one of two non-equivalent direction $[110]$ and $[1\underline{1}0]$ axis correspondingly.
These two species seem to evolve differently under the VBO change, leading to the observed ``levels crossings'' [Fig.~\ref{disc-en} (g) and (h)].
It is important to reiterate at this point that modification of the VBO is ``artificial'', however it constitutes a very interesting theoretical tool.

\subsection{Strained valence band offset}
As expected, InAs/GaAs systems are affected by strain effects more than InAs/InP due to the large lattice mismatch of the former.
For a small value of VBO~$=50$~meV, the confining potential for holes is dominated by the strain contribution and the character (anisotropy) of this term
leads to the ground hole state of the apparent, yet unusual \textit{p}-like symmetry (Fig.~\ref{lens-ho-str}).
I demonstrate results of the calculation for the particularly low ($50$~meV) VBO value as this number is customarily utilized in the empirical pseudopotential method (EPM\cite{Zunger-EPM1,Zunger-EPM2}).
However, I emphasize that it is the strained band offset (being the ``combination'' of both VBO and $a_v$ valence band deformation potential) that ``enters'' the calculation as the actual hole confining potential\cite{zielinski-including}. Whereas the ``natural'' (strain free) VBO and $a_v$ differ significantly between two methods (ETB: $a_v=1$~eV and EPM: $a_v=-1$~eV), the strained band offset obtained by using two sets of $a_v$ and VBO parameters is very similar $\approx 330$~meV\cite{gong-complexes}. This important conclusion is illustrated on Figure~\ref{conf-pot} showing the confining potential profiles calculated for the InAs/GaAs lens type dot using the Bir-Pikus model\cite{pryor-zunger}.
This plot was obtained utilizing two sets of parameters: the ``recommended'' ones from the review paper by Vurgraftman et al.~\cite{VBO-Vurgraftman}
($a_v=+1.0$ eV, VBO~$=0.21$~eV) and those reported\cite{Zunger-EPM1} by the EPM method ($a_v=-1.0$~eV, VBO~$=0.05$~eV).
Thus, in result, the effective confining potentials in the ETB and the EPM approaches are quite similar despite noticeably different bulk (``intermediate'') target values. A more quantitative comparison of the quantum dot results obtained by the ETB and EPM was presented in our recent work\cite{zielinski-including}.

The main difference between the lens type and disk type quantum dots comes from the fact that the disk type quantum dots are subject to ``smooth'', slowly spatially varying strain (Fig.~\ref{trace-biax}),
which also induces uniform heavy hole-light hole splitting leading to single particle-like picture of hole states.
On the contrary, lens type quantum dots are affected by a spatially fluctuating strain due to the curved shape of quantum dots and the presence of jagged, step-like material interface (Fig.~\ref{trace-biax}).
Electron states are built predominately from \textit{s}-type atomic orbitals and are affected only (in the Bir-Pikus formalism\cite{bir-pikus,yu-cardona,pryor-zunger}) by the hydrostatic strain ($Tr(\epsilon)=\epsilon_{xx}+\epsilon_{yy}+\epsilon_{zz}$), that leads predominantly to simple energetic upward shift due to bond lengths contraction in the strained system.
Thus electron states are less affected by highly spatially variable biaxial strain.

On the contrary, heavy-holes confining potential (in the Bir-Pikus formalism) along the growth [001] quantum dot axis going through the dot center, is given as\cite{pryor-zunger}: $E_{HH}=a_vTr\left(\epsilon\right)-bB\left(\epsilon\right)$, where $B\left(\epsilon\right)=\epsilon_{zz}-\left(\epsilon_{xx}+\epsilon_{yy}\right)/2$ is the biaxial component of strain, $a_v$ absolute valence band deformation potential and $b$ is the biaxial strain deformation potential ($b_{InAs}=-1.8$~eV\cite{VBO-Vurgraftman}).
Thus, for hole states, the strain related potential shift is an interplay between the hydrostatic and biaxial strain: the biaxial strain  ($-bB\left(\epsilon\right)>0$) deepens the confinement for holes, whereas the hydrostatic strain makes the confining well in the dot region shallower ($a_vTr\left(\epsilon\right)<0$).


Fig.~\ref{conf-hh-all} shows the heavy-hole (in-plane) confining potential (obtained with bulk parameters from Ref.\cite{VBO-Vurgraftman}) calculated for all quantum dots considered in this paper.
For the lens type InAs/GaAs quantum dot one can notice a non-regular, ``jittered'' spatial dependence that should affect the hole shell structure as discussed earlier in the paper.
These oscillations reach up to $50$~meV and even lead to formation of attracting well for holes at the edge of the quantum dot.
The situation is further complicated at the interface as the biaxial strain changes sign\cite{pryor-zunger} and reverses light-heavy hole ordering, thus affecting the valence band mixing.
The interface related oscillation seems to play a small role for the relatively ``low-strain'' and high VBO value InAs/InP lens type quantum dot.

The complicated character of strain actually makes the InAs/GaAs lens type quantum dots holes spectra so different from other quantum dots.
This peculiarity leads, for example, to the non-Aufbau hole charging pattern and has been confirmed by the experiment\cite{ediger-zunger-aufbau,blokland}, the well-established empirical pseudopotential method and also recent k.p calculations.\cite{kp-schliwa}
It is only for large\cite{VBO-Vurgraftman} InAs/GaAs VBO values ($>0.450$~eV) when hole charge distributions of lens type quantum dot resemble [Fig.~\ref{lens-ho-str}] that of the harmonic oscillator-like states.
However even for unrealistic VBO~$>0.7$~eV the hole \textit{p}-states splitting does not drop below $5$~meV.
It should be expected that this effect will be more pronounced in ``tall'' (high aspect ratio\cite{usman1}) lens, cones or pyramid type quantum dots\cite{kp-schliwa}.

Should accurate (order of meV's) modeling of hole states be important, the choice of the ``natural'' VBO will play an important role.
This is again particularly noticeable for the lens type InAs/GaAs quantum dot [Fig.~\ref{lens-en}].
In the ``recommended'' range of VBO values\cite{VBO-Vurgraftman} ($210-350$~meV), hole $h_1-h_2$ level spacing varies between $9$~meV and $14$~meV and spacings of higher lying level change even more substantially, e.g. $h_3-h_4$ spacing varies from $1$~meV to $7$~meV.

\section{Many-body states}
Strain effects and the valence band offset play a fundamental role for the single particle states in quantum dots.
Next, I calculate many-body properties of the single exciton and several excitonic complexes.

\subsection{Exciton fine structure}
Fig.~\ref{fss} shows the excitonic fine structure\cite{bayer-eh} calculated for different VBO values for the lens and disk type InAs/GaAs and InAs/InP quantum dots.
The energy difference between the two bright excitonic states, the so-called bright exciton splitting (BES), is related to the confining potential anisotropy\cite{bayer-eh}.
The BES is thus larger for curved-shape, highly anisotropic, lens type quantum dots, reaching values varying between $40-60$~meV, whereas for disk type quantum dots the BES does not exceed $20$~meV.
These results are in quantitative agreement with recent EPM calculations\cite{singh-bester-eh}, however as noticed by the EPM researchers\cite{singh-bester-ordering} this method systematically predicts much lower BES values than those reported in the experiment (or in the current ETB work).

For lens type quantum dots the BES varies noticeably with respect to the VBO, whereas it has almost flat VBO dependence for disk-shaped nanosystems.
Although Fig.~\ref{fss} is the result of the full many-body configuration interaction calculation\cite{zielinski-prb09}, including single particle states up to the \textit{d}-shell,
the majority of the BES splitting and the overall trend is very well reproduced (not shown here) by a single electron-hole anisotropic exchange (complex) integral\cite{bayer-eh,bryant-fss,zielinski-prb09,lee-johnsson-prb2001}:
\begin{eqnarray*}
V_{eh}^{b-b} \equiv V_{e_{\uparrow}h_{\Downarrow}e_{\downarrow}h_{\Uparrow}} =\int \int \frac{e_{\uparrow}(\mathbf{r})^*h_{\Downarrow}(\mathbf{r'})^*e_{\downarrow}(\mathbf{r'})h_{\Uparrow}(\mathbf{r})}{\epsilon\left(\mathbf{r},\mathbf{r'}\right)|\mathbf{r}-\mathbf{r'}|}d\mathbf{r} d\mathbf{r}',
\end{eqnarray*}
where $\epsilon\left(\vec{r_1},\vec{r_2}\right)$ is the position-dependent dielectric function, (e) electron and (h) hole are in their ground s-states and arrows correspond to carrier quasi-spins.
$V_{eh}^{b-b}$ (as all other Coulomb matrix elements in this paper) is calculated using Eq.~\ref{coulomb-special}.

$V_{eh}^{b-b}$ is responsible for mixing of two bright excitonic states ($\uparrow\Downarrow$ and $\downarrow\Uparrow$) and therefore leads to the BES.
The BES is thus a two-body effect of the \textit{s}-shell electron and hole spatial anisotropy and as such is not directly correlated with the electron \textit{p}-shell splitting/anisotropy (Fig.~\ref{p-shell}).

The BES is larger for the highly strained lens type InAs/GaAs quantum dot (Fig.~\ref{fss}), as the low symmetry ($C_{2v}$) of the confining potential is related to strain\cite{bester-zunger-prb2005}. 
On the other hand the electron-hole anisotropic exchange interaction is also related to the electron-hole ``overlap'' in $V_{eh}^{b-b}$. Therefore, similar confinement of the electron and hole may lead to larger BES.
Thus, for low strain InAs/InP disk type quantum dot, where the electron and hole charge distributions (``envelopes'') are almost identical (as seen previously on Fig.~\ref{disc-el} and Fig.~\ref{disc-ho-str}),
the BES value ($\approx 18\mu$eV) is about two times larger than for the highly strained InAs/GaAs disk type quantum dot ($\approx 8\mu$eV).
This is in contradiction to recent EPM calculations\cite{zunger-eh}, that predict highly reduced fine-structure splitting in InAs/InP quantum dots.
Finally, should the wetting layer be neglected the BES for the disk type quantum dots would be exactly zero due to high ($D_{2d}$) overall (lattice and shape) quantum dot symmetry\cite{singh-bester-eh}.
To summarize, the BES is a non-trivial function of quantum dot shape, shape and quantum dot/matrix materials.

The dark-bright exchange splitting is determined predominantly by a (real) exchange matrix element which also conserves spin ($V_{eh}^{d-b} \equiv V_{e_{\uparrow}h_{\Uparrow}e_{\uparrow}h_{\Uparrow}}$).
The bright-dark exciton splitting does not vary significantly with the VBO: increased confinement of the hole seems to be assisted by the decreased confinement of the electron,
and the overall electron-hole exchange integral and dark-bright exciton splitting does not vary too much.
Based on similar arguments as for the BES, the bright-dark exciton splitting for disk type quantum dots is generally larger than in lens type quantum dots.
Additionally the bright-dark exciton splitting is larger for cases with stronger electron-hole overlap:
For the disc type quantum dot the BES is larger in the InAs/GaAs system, while for the lens type quantum dot the BES is larger in the InAs/InP system.

Finally, the dark exciton splitting is a pronounced and complicated function of the VBO, especially for InAs/GaAs systems,
varying over an order of magnitude for the studied VBOs.
The dark exciton splitting is generally larger for smaller VBO value, however no clear trend regarding the matrix material is observed.
It should be emphasized that as the dark exciton splitting values are typically on the order of few $\mu$eV's, i.e. $10^{-6}$ smaller than the direct electron-hole Coulomb attraction, any practical calculation of the
dark exciton splitting, aiming for the quantitative agreement with an experiment, should be supported by a careful error analysis.

\subsection{XX, X$^{-}$ and X$^{+}$ binding energies}
Figure~\ref{binding} shows the biexciton (XX) and charged trions (X$^-$ and X$^+$) binding energies, measured with respect to the single exciton energy ($E^{XX}_B=E^{XX}-E^{X}, E^{X^{-}}_B=E^{X^{-}}-E^{X}$, etc.),
for different VBO values, for lens and disk type InAs/GaAs and InAs/InP quantum dots.

For the InAs/GaAs lens type quantum dot, the biexciton binding energy varies significantly and even changes sign at VBO~$\approx 170$~meV, leading to the unbound biexciton for VBO~$<170$~meV.
Similarly, binding energies of charged excitons vary over a large range of values: there is a bound to unbound transition for X$^-$ at VBO~$\approx 0.35$~eV.
For InAs/InP lens type quantum dot, in the ``realistic''\cite{VBO-Vurgraftman,Wei-DFT2,Wei-Zunger-DFT,VanDeWalle,kend-hart-zunger,kadantsev-ziel-kork-haw,kadantsev,kadantsev-ziel-haw} range of InAs/InP VBO values ($300-400$~meV), charged complexes even reverse their relative position in a non-trivial pattern, e.g. for VBO~$=0.3$~eV, one observes the following ordering of spectral lines with the increasing energy: XX, X$^+$, X$^-$, X, whereas for VBO~$\approx0.4$~eV the order of lines is following: X$^-$, XX, X$^+$, X.

For the InAs/GaAs disk type quantum dot (``realistic''\cite{VBO-Vurgraftman,Wei-DFT2,Wei-Zunger-DFT,VanDeWalle,kend-hart-zunger,kadantsev-ziel-kork-haw,kadantsev,kadantsev-ziel-haw} VBO~$\approx 200-300$~meV), the following order of spectral lines is observed: X$^-$, XX, X$^+$ and X, whereas for the InAs/InP disk type quantum dot (VBO~$\approx 300-400$~meV) I obtain:  X$^-$, XX, X and X$^+$.
In all considered cases absolute binding energies and the relative order of spectral lines depend significantly on the VBO, quantum dot shape and substrate material.
Therefore applying different VBO values in the excitonic calculation should be supported be a careful VBO analysis.

Excitonic binding energies must be calculated using the full many-body approach\cite{wojs-hawrylak,gong-complexes,zielinski-substrate}:
\begin{eqnarray*}
\label{HFlevel}
E^{XX}_B&=&J_{ss}^{ee}+J_{ss}^{hh}-2J_{ss}^{eh}-\Delta E_{corr}^{XX}\\
E^{X^{-}}_B&=&J_{ss}^{ee}-J_{ss}^{eh}-\Delta E_{corr}^{X^-}\\
E^{X^{+}}_B&=&J_{ss}^{hh}-J_{ss}^{eh}-\Delta E_{corr}^{X^+}
\end{eqnarray*}
where electron-electron ($J_{ee}\equiv V_{e_1h_1e_1h_1}$), hole-hole ($J_{hh}\equiv V_{h_1h_1h_1h_1}$) and electron-hole $J_{eh}\equiv V_{e_1h_1h_1e_1}$ Coulomb integrals are calculated for the electron and the hole occupying their (\textit{s}-shell) ground states ($e_1$ and $h_1$), whereas the important correction due to correlation effects $\Delta E_{corr}$ can be attributed to the configuration mixing effects with higher lying states.
In the Hartree-Fock (perturbation theory) approximation $\Delta E_{corr}=0$, whereas realistic values of binding energies ($\Delta E_{corr}\neq0$) can be calculated using the full configuration interaction method\cite{wojs-hawrylak,gong-complexes,zielinski-substrate}.

For example, for the lens type InAs/GaAs quantum dot, at the VBO$=0.1$~eV, I obtain $J_{ee}=26.28$~meV and $J_{eh}=19.54$~meV 
and at the level of Hartree-Fock approximation $X^{-1}$ binding energy can be calculated as $\Delta E^{HF}\left(X^{-}\right)=J_{ss}^{ee}-J_{ss}^{eh}=6.75$~meV.
This value is further reduced by the correlation correction $\Delta E_{corr}^{X^{-}}=1.41$~meV and therefore finally $E^{X^{-}}_B=5.34$~meV.
On the other hand, for VBO$=0.5$~meV I obtain $J_{ee}=20.88$~meV and $J_{eh}=21.68$~meV, leading to $\Delta E^{HF}\left(X^{-}\right)=-0.8$~meV.
The CI calculated correlation correction is $\Delta E_{corr}^{X^{-}}=1.52$~meV and thus we obtain $E^{X^{-}}_B=-2.32$~meV.
Therefore the negatively charged exciton binding energy varies from $5.34$ to $-2.32$~meV for VBO$=0.1$~eV and  VBO$=0.5$~eV correspondingly.
The correction due to correlation mixing is important ($\approx 1.5$~meV) and cannot be neglected, however its value does change significantly as the VBO is varied.
In the previous section I discussed that whereas the absolute values of excitonic fine structure splittings were results of the full configuration interaction procedure,
observed trends could be analyzed in terms of a single exchange integral.
A quite similar situation occurs for excitonic complexes binding energies.
The absolute value of the binding energy must be calculated using the many body approach, yet its evolution with respect to the VBO can be understood at the level of Hartree-Fock (perturbation theory) approximation
and several \textit{s}-shell electron-hole Coulomb integrals ($J_{ee}$, $J_{eh}$ and $J_{hh}$), whereas the correlation correction $\Delta E_{corr}$ is virtually unaffected by the choice of the VBO.

\section{Summary}
Valence band offset is one of the important empirical parameters for the semi-empirical tight-binding method.
By a thorough VBO analysis I have shown a non-trivial interplay between the confinement potential due to the band offset and the confinement due to strain.
With strain effects artificially neglected, no shell structure for holes is present, with complicated charge density distribution due to light and heavy hole band mixing effects.
With strain effects included, heavy holes and light holes are decoupled from each other by the biaxial strain and the shell-like structure characteristic for single band models is restored.
A notable difference from this picture is observed for the strained InAs/GaAs lens type quantum dot.
Due to anisotropy and inhomogeneity of the strained affected confining potential the splittings of excited hole states dominate over shell spacings, even for the largest considered VBO value.
The peculiar structure of hole levels in InAs/GaAs in lens type quantum dots, so much different from InAs/InP quantum dots,
origins thus from strain and is not due to the different VBO as suggested by the other studies\cite{gong-prb77}.
On the other hand the characteristic shell structure of InAs/GaAs disk type quantum dots or InAs/InP lens type quantum dots also origins from strain.
In all considered cases, even for ``low-strain'' InAs/InP systems, strain effects and large VBO value lead to spectral structure of hole levels known from the experiment.
Therefore, strain cannot be neglected even in nominally ``low-strain'' InAs/InP quantum dots, whether disc- or lens-shaped.

Valence band offset also affects many body properties such as excitonic fine structure and binding energies of excitonic complexes.
Exciton fine structure splittings reported in this paper are much larger than those reported by the EPM and
thus much closer to the experimentally reported values without the necessity of the inclusion of the ``ordering'' effects\cite{singh-bester-ordering}.
Charged excitons and biexciton binding energies depend significantly on the depth of confinement potential (VBO) with clear distinction from the simple effective mass (``harmonic oscillator'') picture.
This should be emphasized as the reliable prediction of excitonic binding energies is of key importance for so-called ``inverse approaches''\cite{mlinar-loop}.

The paper studies the differences between flat (``indium flushed'') and lens type quantum dots, the relation between hydrostatic and biaxial strain in these systems, and quantitatively estimates the role of a jagged quantum dot/matrix material interface. The paper shows that the unrealistic \textit{p}-shell symmetry of ground hole state would be obtained if the ETB used the VBO value taken directly from the EPM and then discusses differences and similarities (``strained band offset'') between both approaches.
Finally the paper shows that the VBO is not merely a technological parameter, and the caution should be exercised when applying different VBO values.

\section{Acknowledgements}
This work was supported by the Foundation for Polish Science, Homing Plus Programme co-financed by the European Union within the European
Regional Development Fund.
The author would like to thank G. W. Bryant, W. Jask\'olski, and E. Kadantsev for discussions and patient reading of the manuscript.

\newpage









\begin{widetext}

\begin{figure}
  \begin{center}
  \includegraphics[width=0.8\textwidth]{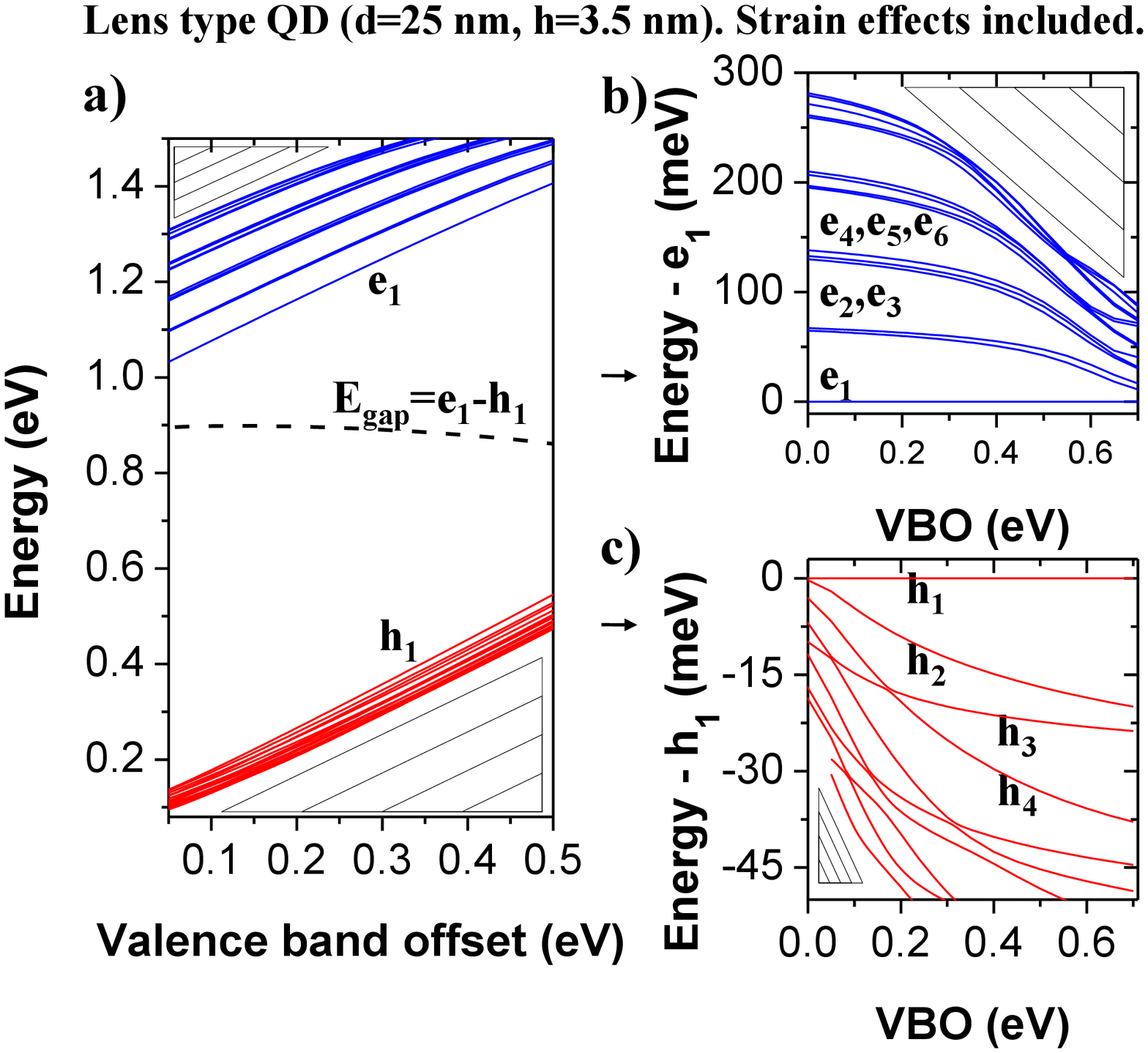}
  \end{center}
  \caption{
  (Color online) (a) Single particle (electron and hole) energies and the effective gap $E_{gap}=e_1-h_1$ for InAs/GaAs lens type (d=$25$ nm, h=$3.5$ nm) quantum dot as a function of quantum dot (InAs) and matrix (GaAs) valence band offset (VBO), (b) single particle electron and (c) hole energies are calculated with respect to the electron ($e_1$) and hole ($h_1$) ground state energies.
  Patterned areas mark higher, excited states. }
  \label{lens-example}
\end{figure}

\begin{figure}
  \begin{center}
  \includegraphics[width=0.8\textwidth]{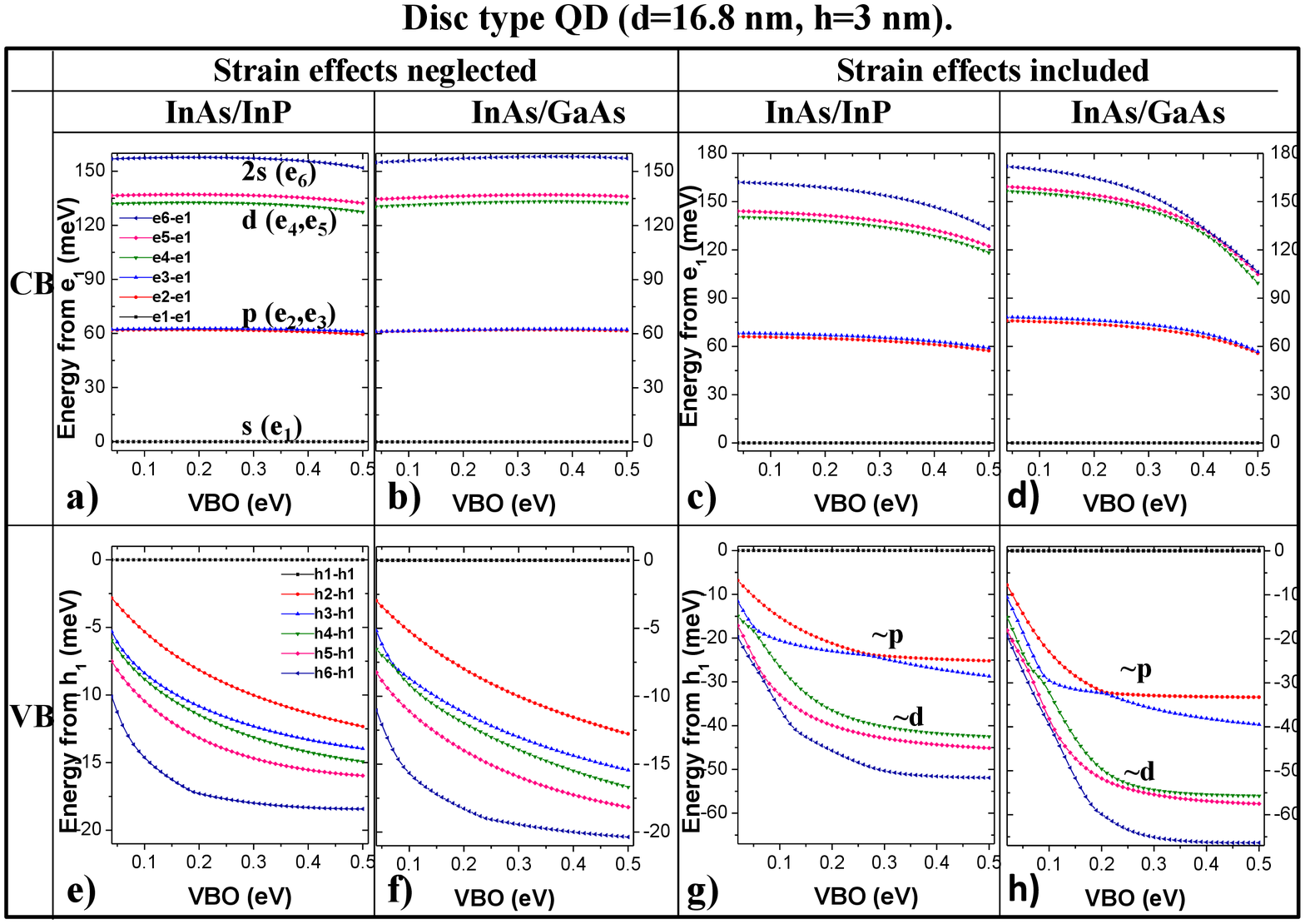}
  \end{center}
  \caption{
  Electron (conduction band - CB) and hole (valence band - VB) single particle energies calculated with respect to the electron and hole ground state energies ($e_1$ and $h_1$ correspondingly) for InAs/InP and InAs/GaAs disk type (d=$16.8$ nm, h=$3$ nm) quantum dots as a function of quantum dot (InAs) and matrix (GaAs or InP) valence band offset (VBO). Strain-effects are either included or artificially neglected.}
  \label{disc-en}
\end{figure}

\begin{figure}
  \begin{center}
  \includegraphics[width=0.8\textwidth]{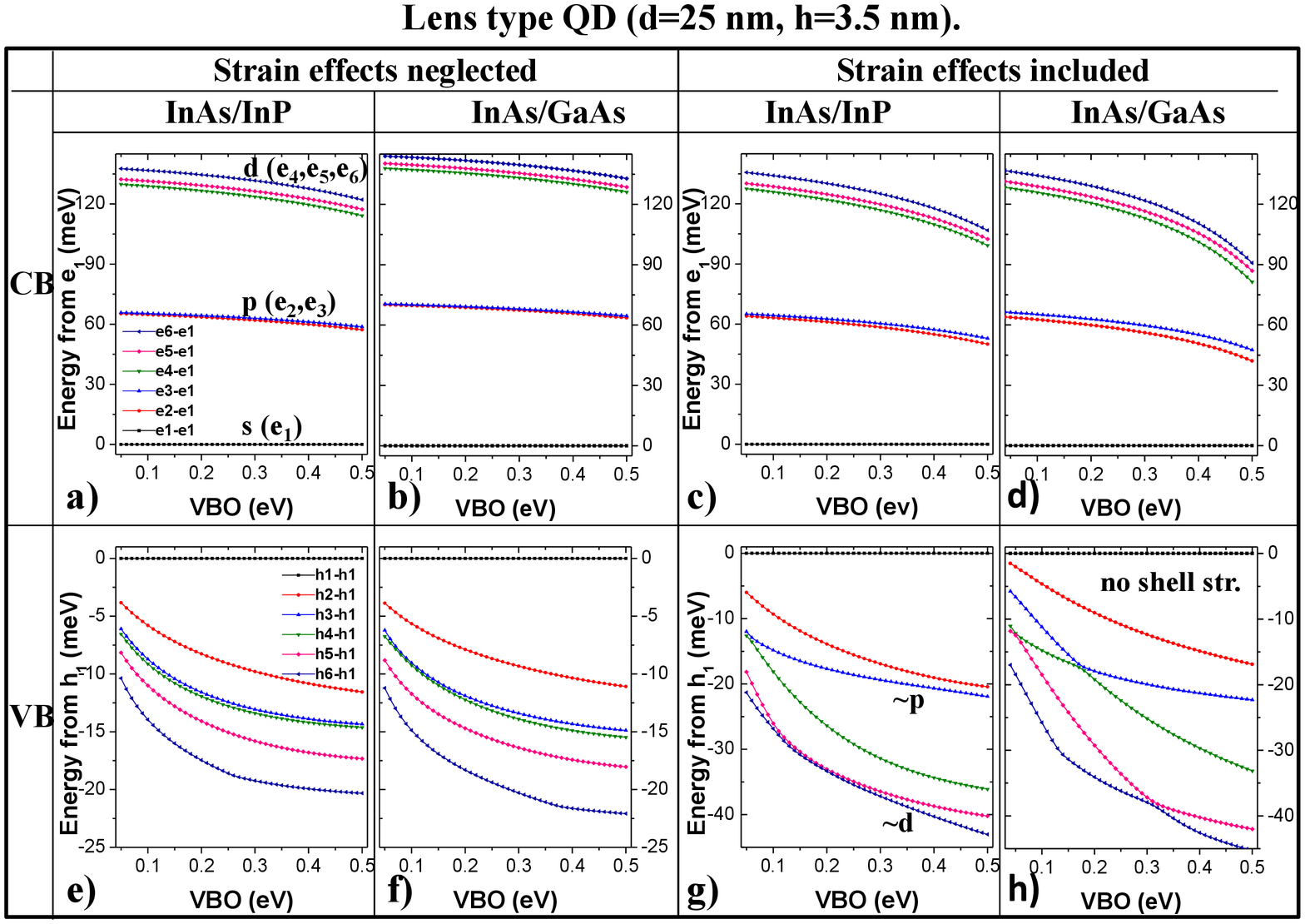}
  \end{center}
  \caption{
  Electron (conduction band - CB) and hole (valence band - VB) single particle energies calculated with respect to the electron and hole ground state energies ($e_1$ and $h_1$ correspondingly) for InAs/InP and InAs/GaAs lens type (d=$25$ nm, h=$3.5$ nm) quantum dots as a function of quantum dot (InAs) and matrix (GaAs or InP) valence band offset (VBO). Strain-effects are either included or artificially neglected.}
  \label{lens-en}
\end{figure}

\begin{figure}
  \begin{center}
  \includegraphics[width=0.8\textwidth]{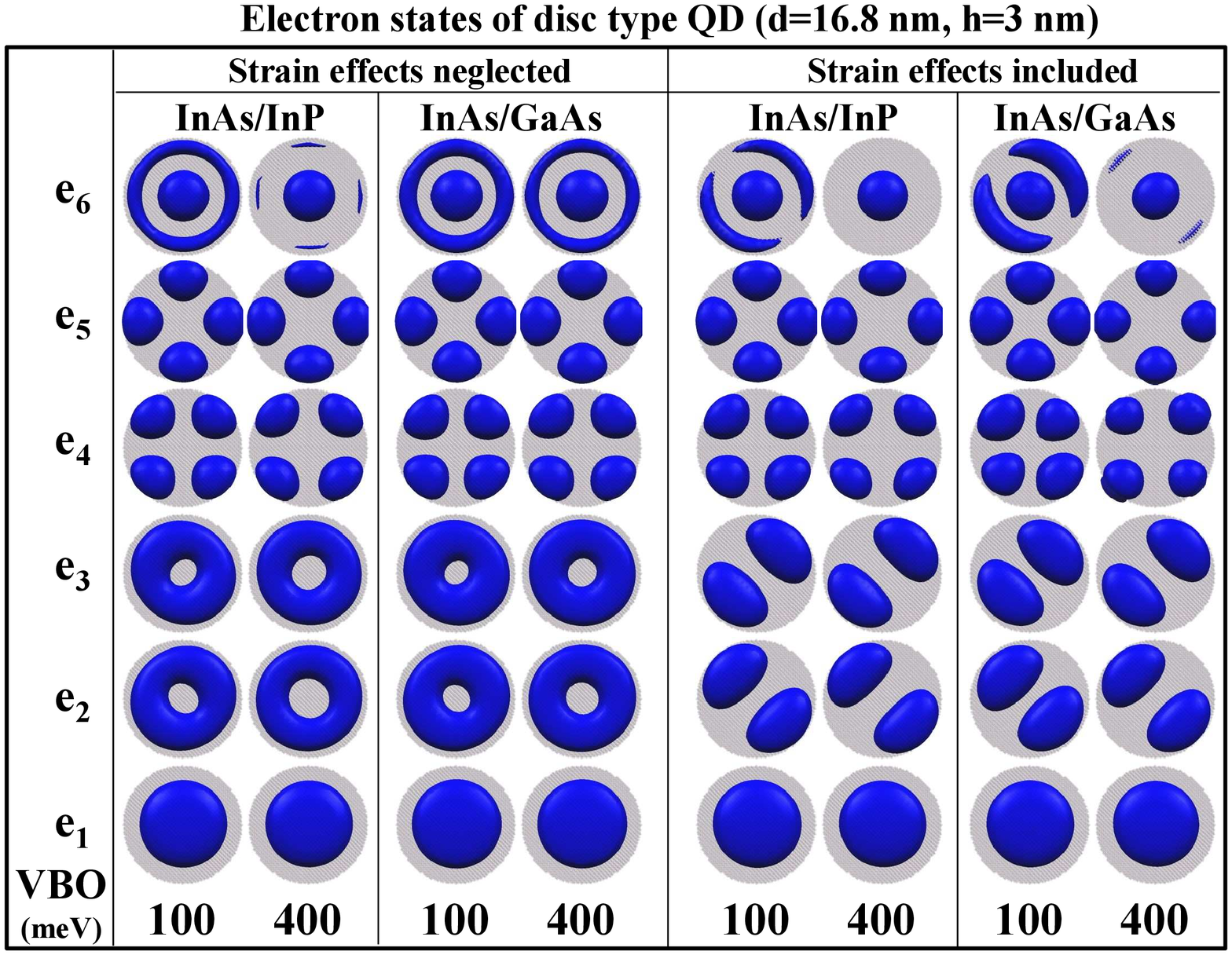}
  \end{center}
  \caption{
  Electron probability density isosurfaces for InAs/InP and InAs/GaAs disk type (d=$16.8$ nm, h=$3$ nm) quantum dots as a function of quantum dot (InAs) and matrix (GaAs or InP) valence band offset (VBO).
  Strain-effects are either included or artificially neglected.}
  \label{disc-el}
\end{figure}

\begin{figure}
  \begin{center}
  \includegraphics[width=0.8\textwidth]{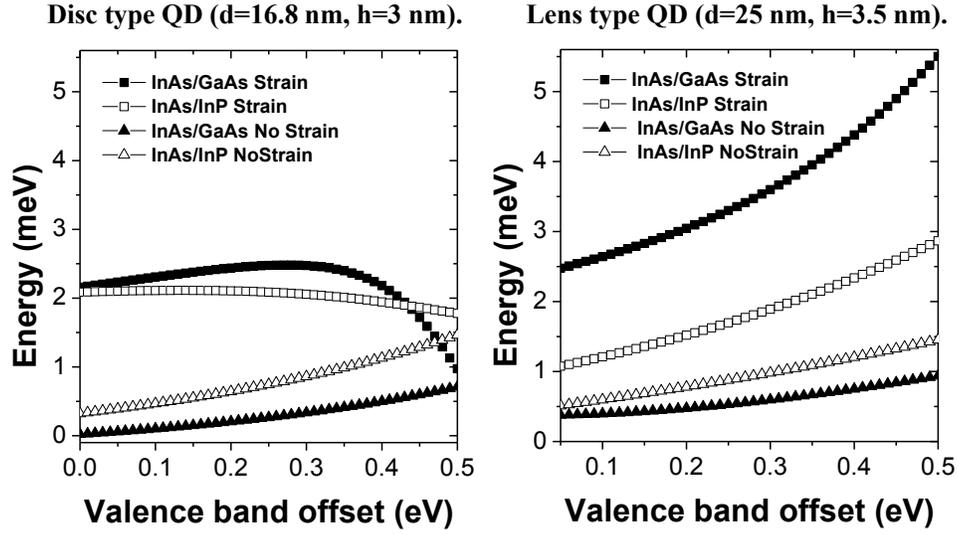}
  \end{center}
  \caption{
  Splitting of the electronic \textit{p}-shell calculated for InAs quantum dots of different shape, matrix material and as a function of quantum dot (InAs) and matrix (GaAs or InP) valence band offset (VBO).}
  \label{p-shell}
\end{figure}

\begin{figure}
  \begin{center}
  \includegraphics[width=0.8\textwidth]{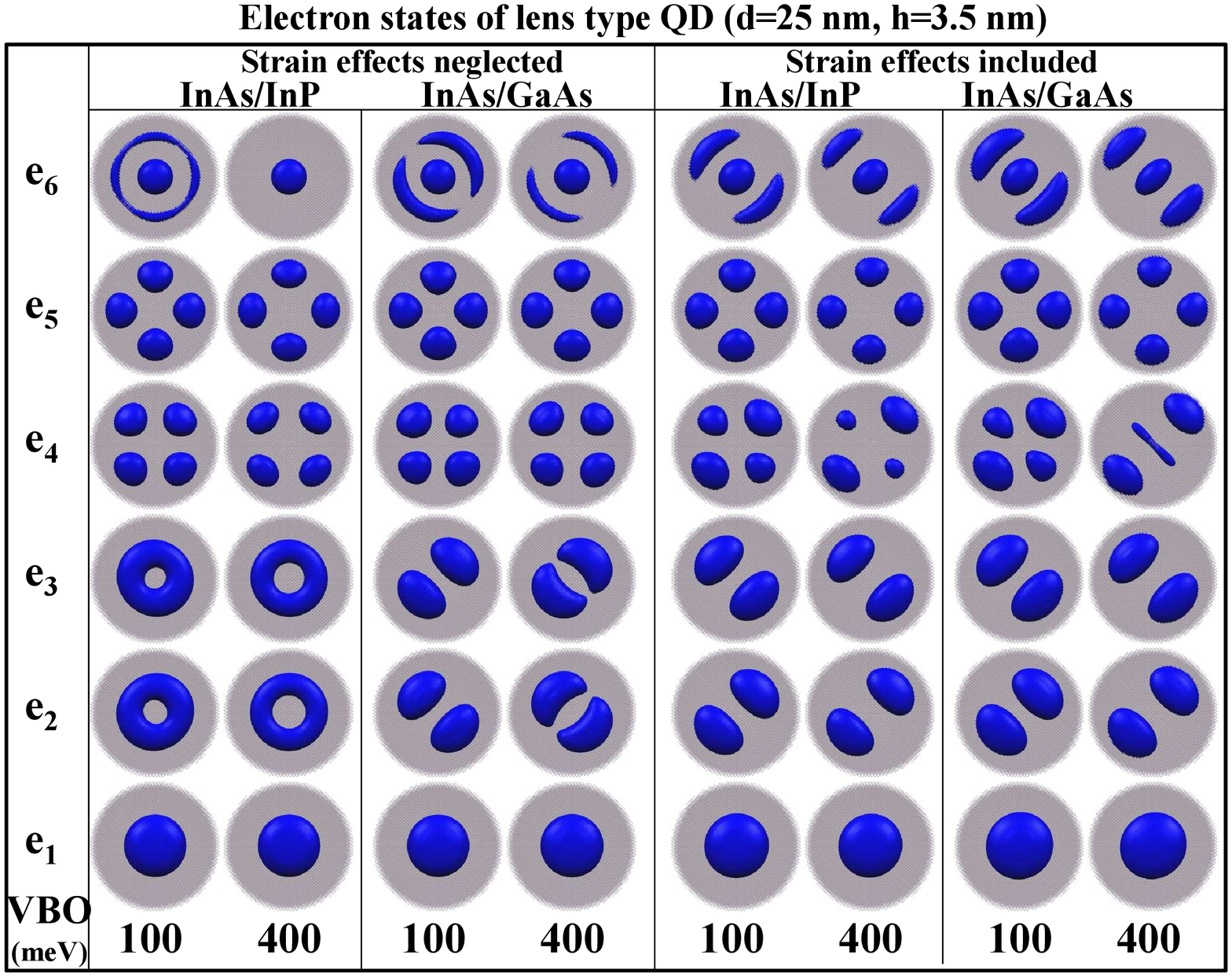}
  \end{center}
  \caption{
  Electron probability density isosurfaces in InAs/InP and InAs/GaAs lens type (d=$25$ nm, h=$3.5$ nm) quantum dots as a function of quantum dot (InAs) and matrix (GaAs or InP) valence band offset (VBO).
  Strain-effects are either included or artificially neglected.}
  \label{lens-el}
\end{figure}

\begin{figure}
  \begin{center}
  \includegraphics[width=0.8\textwidth]{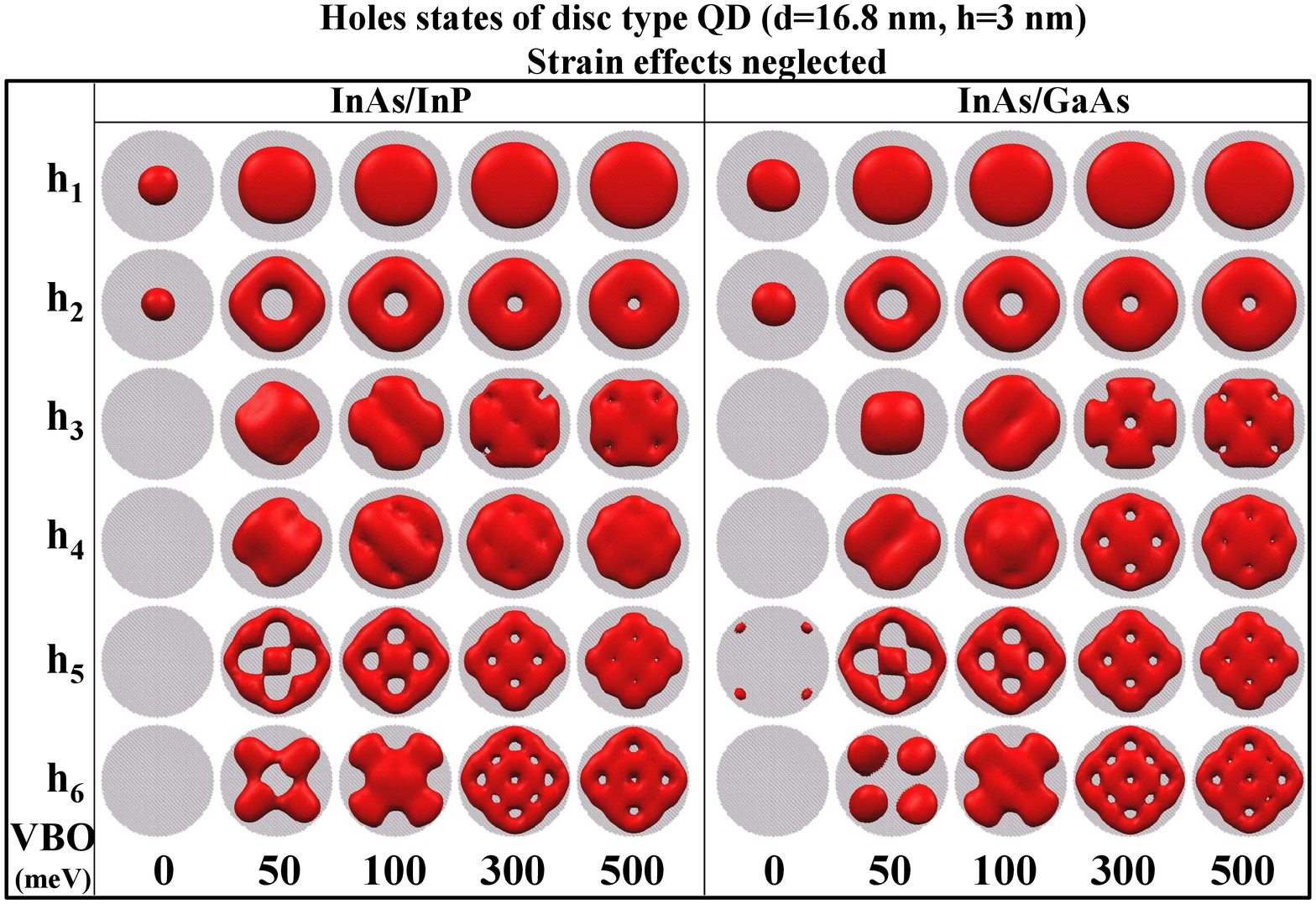}
  \end{center}
  \caption{
  Hole probability density isosurfaces in InAs/InP and InAs/GaAs disk type (d=$16.8$ nm, h=$3$ nm) quantum dots as a function of quantum dot (InAs) and matrix (GaAs or InP) valence band offset (VBO).
  Strain-effects are artificially neglected.}
  \label{disc-ho-nstr}
\end{figure}

\begin{figure}
  \begin{center}
  \includegraphics[width=0.8\textwidth]{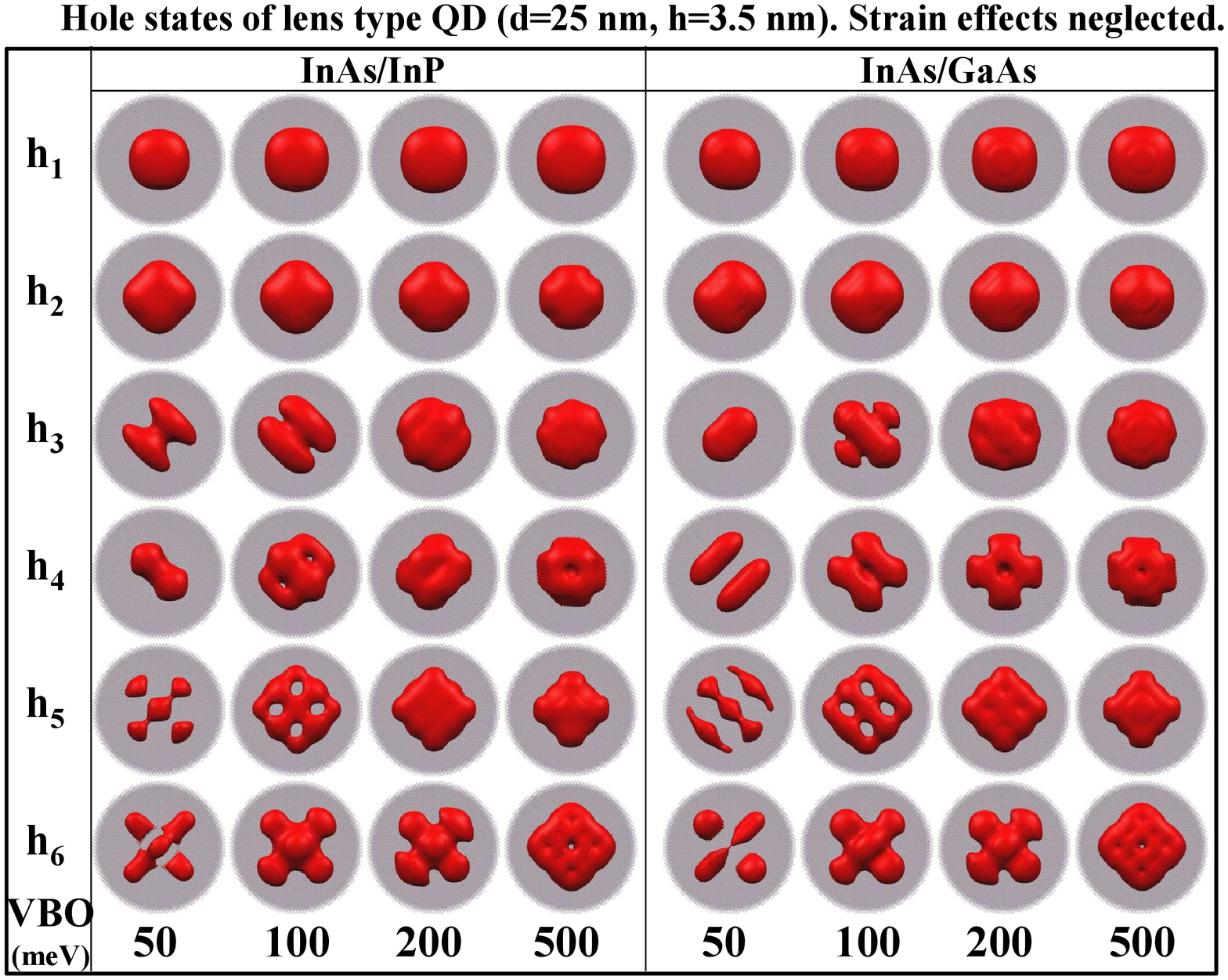}
  \end{center}
  \caption{
  Hole probability density isosurfaces in InAs/InP and InAs/GaAs lens type (d=$25$ nm, h=$3.5$ nm) quantum dots as a function of quantum dot (InAs) and matrix (GaAs or InP) valence band offset (VBO).
  Strain-effects are artificially neglected.}
  \label{lens-ho-nstr}
\end{figure}

\begin{figure}
  \begin{center}
  \includegraphics[width=0.8\textwidth]{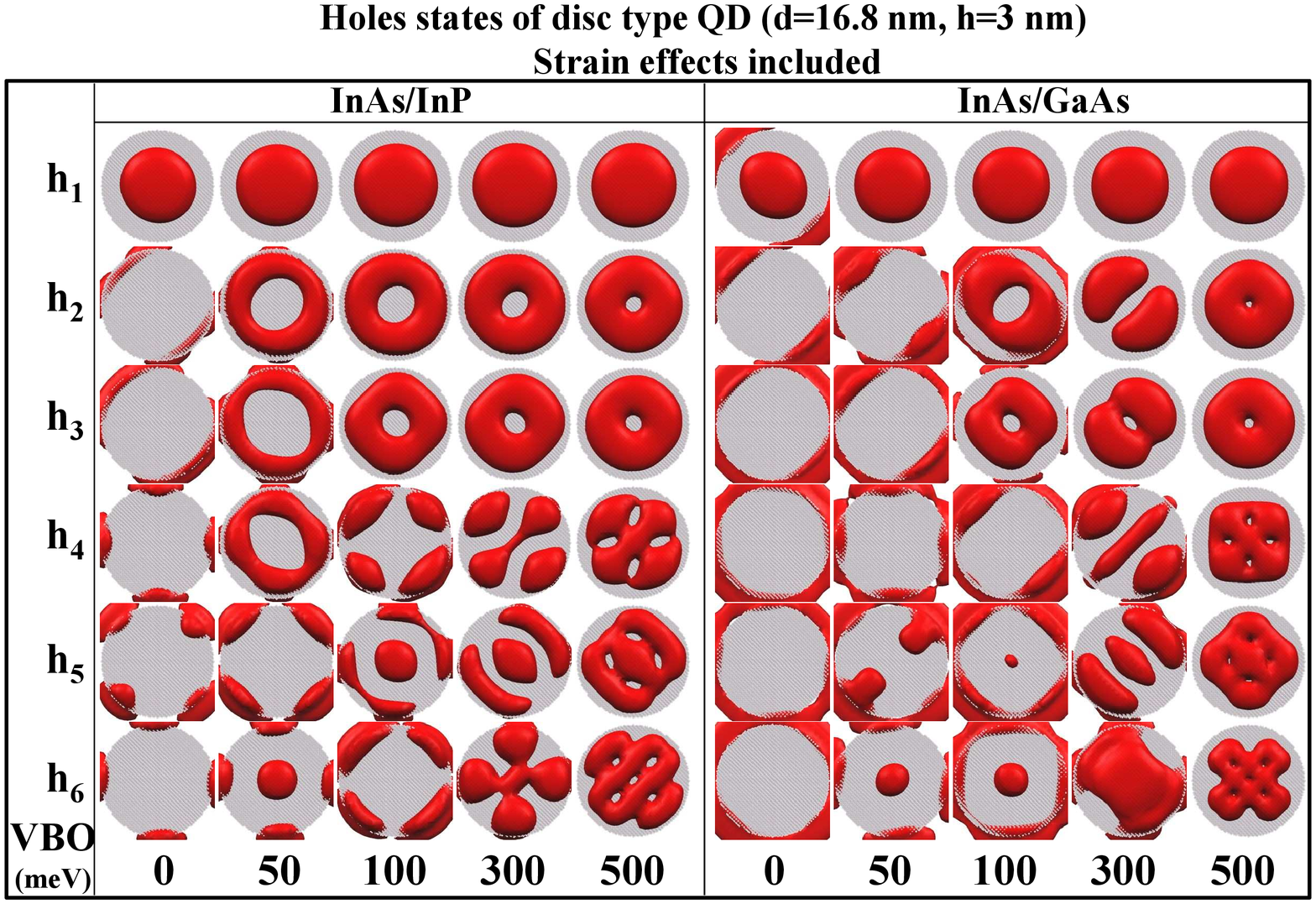}
  \end{center}
  \caption{
  Hole probability density isosurfaces in InAs/InP and InAs/GaAs disk type (d=$16.8$ nm, h=$3$ nm) quantum dots as a function of quantum dot (InAs) and matrix (GaAs or InP) valence band offset (VBO).
  Strain-effects are included.}
  \label{disc-ho-str}
\end{figure}

\begin{figure}
  \begin{center}
  \includegraphics[width=0.8\textwidth]{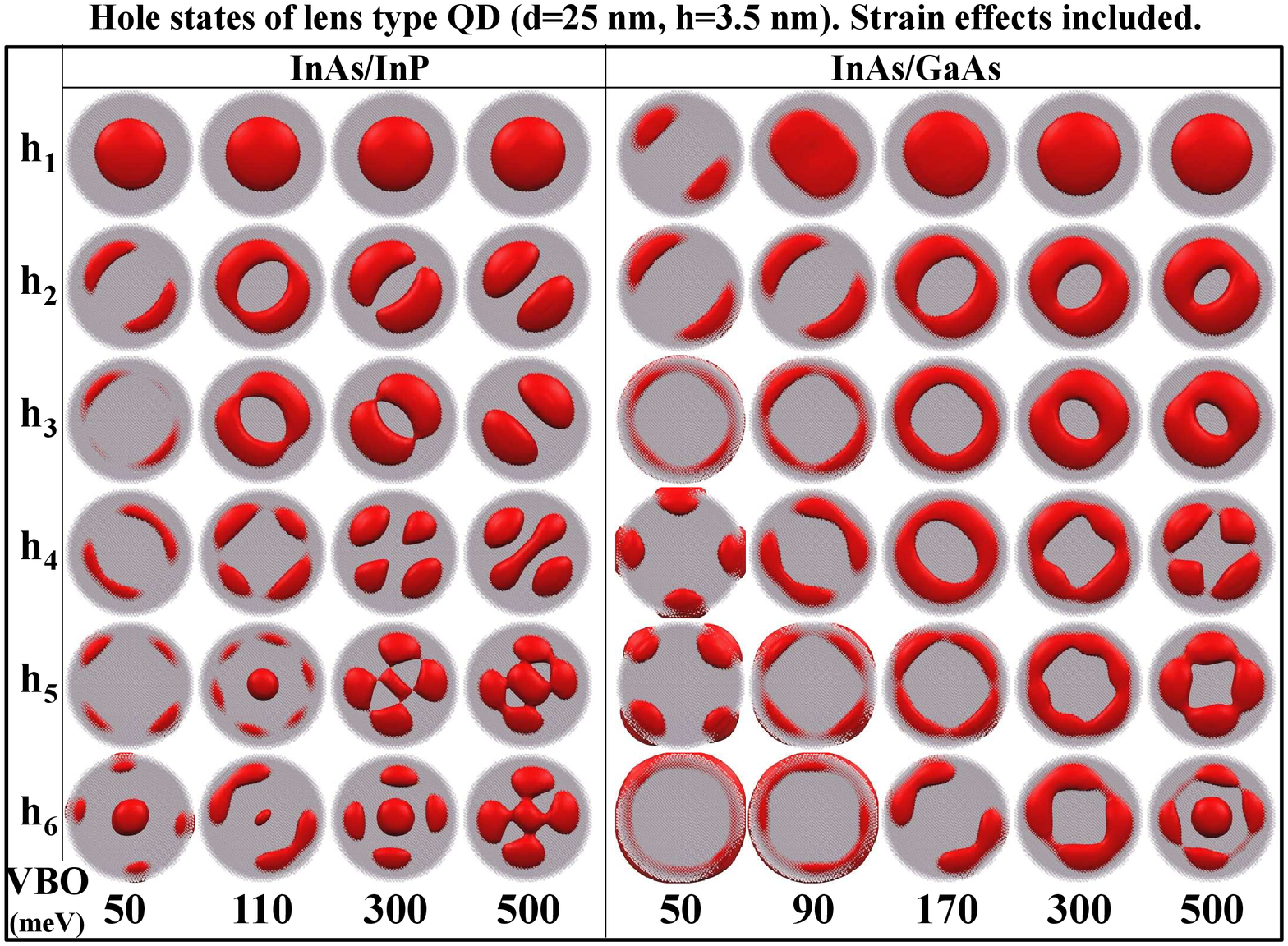}
  \end{center}
  \caption{
  Hole probability density isosurfaces in InAs/InP and InAs/GaAs lens type (d=$25$ nm, h=$3.5$ nm) quantum dots as a function of quantum dot (InAs) and matrix (GaAs or InP) valence band offset (VBO).
  Strain-effects are included.}
  \label{lens-ho-str}
\end{figure}

\begin{figure}
  \begin{center}
  \includegraphics[width=0.8\textwidth]{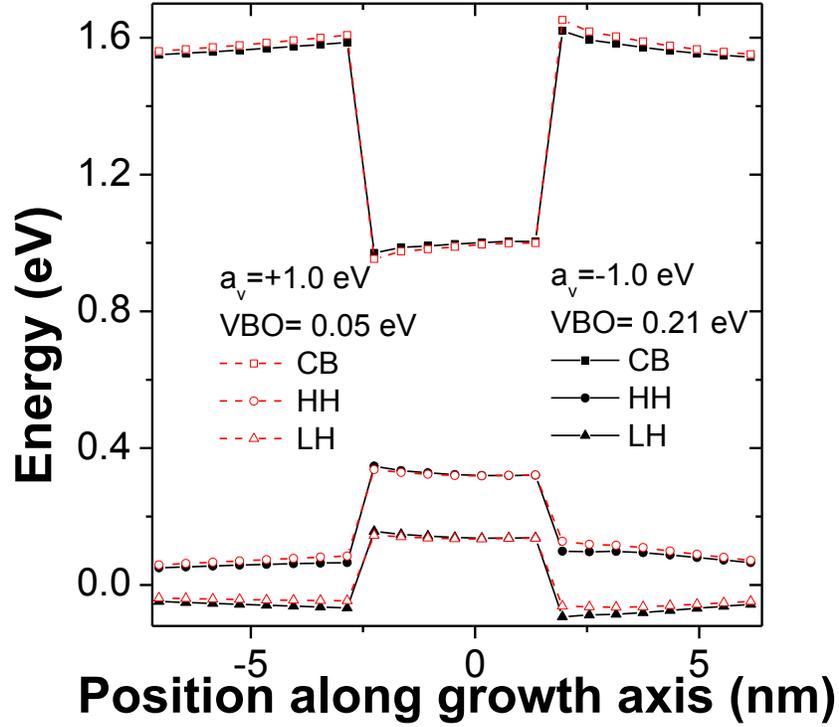}
  \end{center}
  \caption{
  Strain-induced confining potentials for a InAs/GaAs lens type (d=$25$ nm, h=$3.5$ nm) quantum dot along [001] axis, calculated using the Bir-Pikus model and two distinct sets of bulk VBO and $a_v$ parameters.
  Conduction band (CB) - squares, heavy-hole band (HH) - circles, light-hole band (LH) - triangles.}
  \label{conf-pot}
\end{figure}

\begin{figure}
  \begin{center}
  \includegraphics[width=0.8\textwidth]{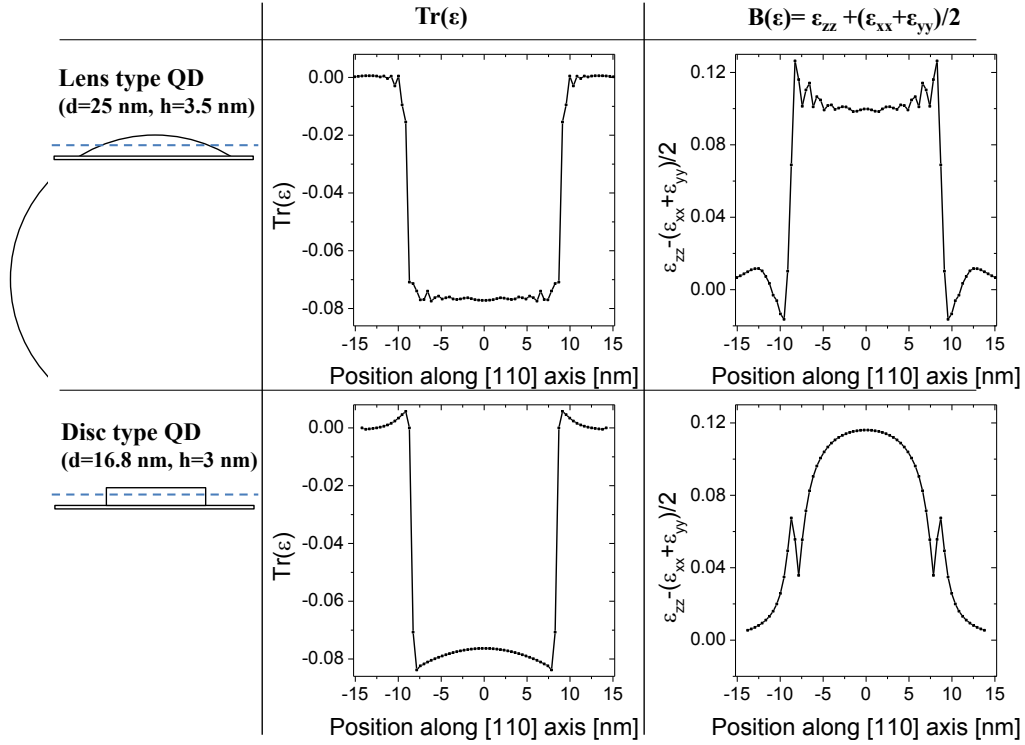}
  \end{center}
  \caption{
  Trace $Tr\left(\epsilon\right)$ of the strain tensor and the biaxial $B\left(\epsilon\right)=\epsilon_{zz}-\left(\epsilon_{xx}+\epsilon_{yy}\right)/2$ component of strain for a disk type (d=$16.8$ nm, h=$3$ nm) and a lens type (d=$25$ nm, h=$3.5$ nm) InAs/GaAs quantum dot. Profiles have been calculated along [110] direction, $z=2$~nm from the quantum dot base.}
  \label{trace-biax}
\end{figure}

\begin{figure}
  \begin{center}
  \includegraphics[width=0.8\textwidth]{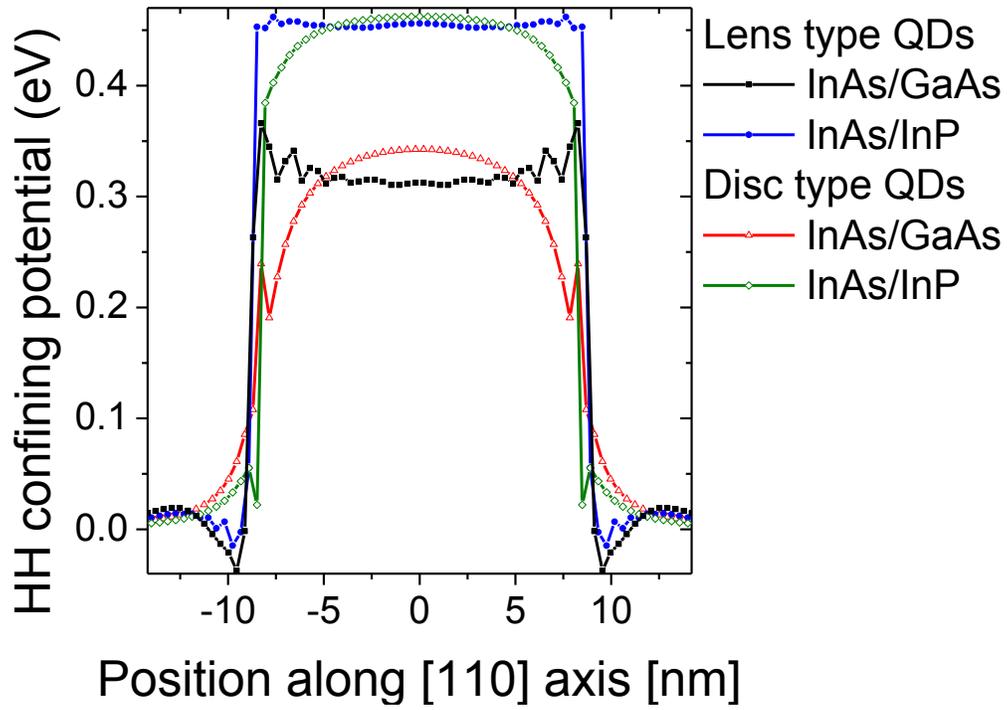}
  \end{center}
  \caption{
  Strain-induced heavy-hole confining potentials for InAs/GaAs and InAs/InP disk type (d=$16.8$ nm, h=$3$ nm) and lens type (d=$25$ nm, h=$3.5$ nm) quantum dots.
  Profiles has been calculated using the Bir-Pikus model along [110] direction, $z=2$~nm from the quantum dot base.}
  \label{conf-hh-all}
\end{figure}

\begin{figure}
  \begin{center}
  \includegraphics[width=0.8\textwidth]{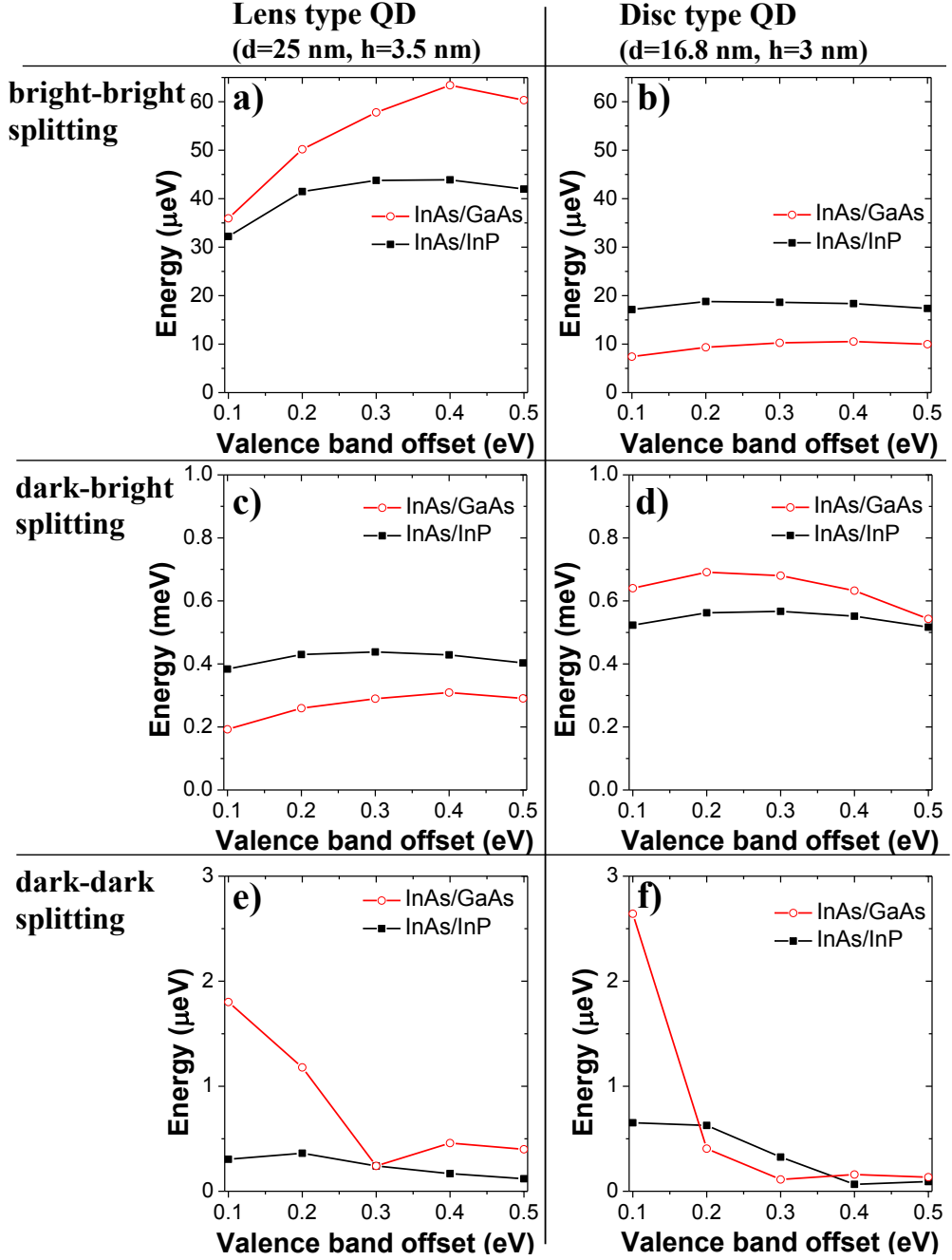}
  \end{center}
  \caption{
  Excitonic fine structure for InAs/GaAs (open circles) and InAs/InP (squares) disk type (d=$16.8$ nm, h=$3$ nm) and lens type (d=$25$ nm, h=$3.5$ nm) quantum dots as a function of quantum dot (InAs) and matrix (GaAs or InP) valence band offset (VBO).}
  \label{fss}
\end{figure}

\begin{figure}
  \begin{center}
  \includegraphics[width=0.8\textwidth]{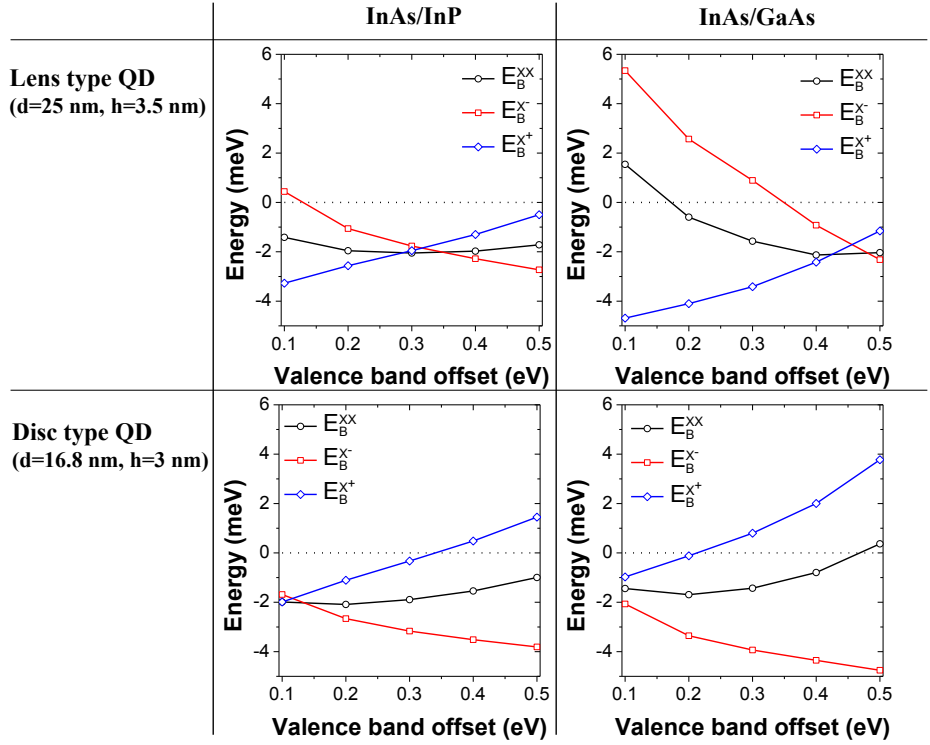}
  \end{center}
  \caption{
  Biexciton XX (black-circles), negatively charged X$^-$ (red-squares) and positively charged X$^+$ (blue-diamonds) excitons binding energies, calculated with respect to the single, neutral, exciton energy, for InAs/GaAs and InAs/InP disk type (d=$16.8$ nm, h=$3$ nm) and lens type (d=$25$ nm, h=$3.5$ nm) quantum dots as a function of quantum dot (InAs) and matrix (GaAs or InP) valence band offset (VBO).}
  \label{binding}
\end{figure}

\end{widetext}

\end{document}